\documentclass[prc,twoside,superscriptaddress,showkeys,showpacs,twocolumn]{revtex4}

\usepackage{graphicx, latexsym, amssymb, amsmath, color, multirow, mathrsfs, CJK, ifpdf}
\usepackage[section]{placeins}
\usepackage{amsmath}
\usepackage{amsfonts}
\usepackage{amssymb}
\usepackage{bm}% bold math
\usepackage{graphicx}
\usepackage{color} %支持彩色字体
\usepackage{txfonts}
\newcommand{\lrb}[1]{\left(#1\right)}
\newcommand{\lrlc}[1]{\left|#1\right>}

\newcommand{\lrs}[1]{\left[#1\right]}

\newcommand{\svec}[1]{\boldsymbol{ #1}}

\usepackage{ulem}

\newcommand{\delete}{\bgroup\markoverwith{\textcolor{red}{\rule[0.5ex]{2pt}{1pt}}}\ULon}

\renewcommand{\emph}[1]{{\bfseries#1}}

\makeatletter

\newcommand{\Rmnum}[1]{\expandafter\@slowromancap\romannumeral#1@}
\makeatother

\begin{document}

\preprint{APS/123-QED} 
\title{Odd-even staggerings on nuclear binding energy described by the covariant density functional theory}
\author{Long Jun Wang }
\affiliation{School of Nuclear Science and Technology, Lanzhou University, Lanzhou 730000, China}
\author{Bao Yuan Sun }
\affiliation{School of Nuclear Science and Technology, Lanzhou University, Lanzhou 730000, China}
\author{Jian Min Dong }
\affiliation{Institute of Modern Physics, Chinese Academy of Science, Lanzhou 730000, China}
\author{Wen Hui Long }
\email{longwh@lzu.edu.cn}
\affiliation{School of Nuclear Science and Technology, Lanzhou University, Lanzhou 730000, China}

\date{\today}
%%%%%%%%%%%%%%%%%%%%%%%%%%%%%%%%%%%%%%%%%%%%%%%%%%%%%%%%%%%%%%%%%%%%%%%%%%%%%%%%%%%%%%%%%%%%%%%%%%%%%%%%%%%%%%%%%%%%%%%%%%%%%%%%%%%%%%%%%%%%%%%%
\begin{abstract}
The odd-even staggerings (OES) on nuclear binding energies are studied systematically within the covariant density functional (CDF) theories, specifically the relativistic Hartree-Fock-Bogoliubov (RHFB) and the relativistic Hartree-Bogoliubov (RHB) theories. Taking the finite-range Gogny force D1S as an effective pairing interaction, both CDF models can provide appropriate descriptions on the OESs of nuclear binding energies for C, O, Ca, Ni, Zr, Sn, Ce, Gd and Pb isotopes as well as for $N=50$ and $82$ isotones. However, due to the inconsistence between the non-relativistic pairing interaction and the relativistic effective Lagrangians, there exist some systematical discrepancies from the data, i.e., the underestimated OESs in light C and O isotopes and the overestimated ones in heavy region, respectively. Such discrepancies can be eliminated partially by introducing a $Z$- or $N$-dependent strength factor into the pairing force Gogny D1S. In addition, successful descriptions of the occupation numbers of Sn isotopes are achieved with the optimized Gogny pairing force. Furthermore, the analysis of the systematics of both pairing effects and nuclear binding energy indicate the requirement of an unified relativistic mechanism in both p-p and p-h channels to improve the quantitative precision of the theory.
\end{abstract}

\maketitle
 
\section{INTRODUCTION}\label{intro}

In nuclear physics, the pairing mechanism is one of the basic ingredients in determining the nuclear structure properties \cite{Bohr1998nuclear}, and becomes even more significant for the exotic and superheavy nuclei which have attracted wide interests of the community during the past decades \cite{Vretenar1997PRC, Lalazissis1998PRC, Lalazissis1998NPA, Meng1996PRL, Meng1998PRL, Hagino2007PRL, Sun2010PLB, Sun2012PRC, Sobiczewski2007PPNP, Karatzikos2010PLB, Lalazissis1996NPA}. Specifically in exotic nuclei \cite{ExoticReview2000PPNP}, the weakly bound nuclear systems under extreme condition, crucial roles played by the pairing correlations are found not only in predicting the isospin limits of finite nuclei \cite{Vretenar1997PRC, Lalazissis1998PRC, Lalazissis1998NPA}, but also in developing the exotic modes, such as nuclear halo structures and possible BCS-BEC crossover \cite{Meng1996PRL, Meng1998PRL, Hagino2007PRL, Sun2010PLB, Sun2012PRC}. For superheavy nuclei with extra large charge numbers, not only the detailed microscopic structure, but also the bulk properties, e.g., nuclear shapes, fission barriers and collective modes, are essentially related with the effects of pairing \cite{Sobiczewski2007PPNP, Karatzikos2010PLB, Lalazissis1996NPA}.

While experimentally it is not so straightforward to measure the pairing effects, which are in general evaluated by the odd-even staggering (OES) on the nuclear binding energy. The binding energy of a system with odd nucleon (neutron or proton) number is found to be lower than the arithmetic mean of two even neighbors, which leads to the so-called OES of single-nucleon separation energies along the isotopic or isotonic chains. Since the early days of nuclear physics, the OES was interpreted as the presence of pairing correlations between nucleons in nucleus \cite{Bohr1998nuclear}. In the independent quasi-particle picture, the OES extracted from the experimental binding energies is often taken as the reference of the pairing-gap energy, which also provides a quantitative observable to constrain the pairing interaction \cite{Litvinov2005PRL}.

During the past decades, many successes have been achieved by the covariant density functional (CDF) theories in describing the structure properties of nuclear systems \cite{Reinhard1989, Ring1996PPNP, Bender2003RMP, Vretenar2005PhysRep, Meng2006PPNP, Long2006PLB1, Long2007PRC}. One of the most popular CDF models is the relativistic Hartree approach with the no-sea approximation, namely, the relativistic mean field (RMF) theory \cite{Walecka:1974, Serot:1986}, which has been widely applied in exploring the properties of both ground states \cite{Ring1996PPNP, Vretenar2005PhysRep, Meng2006PPNP} and excited states \cite{Vretenar2005PhysRep, Paar2007RPP} for the finite nuclei in and far from the valley of $\beta$ stability. However, due to the limit of the approach itself, significant system degrees of freedom are missing in RMF, such as the one-pion exchange \cite{Lalazissis2009PRCR} and $\rho$-tensor couplings. With the growth of computational facilities and the development of new methods, such defects can be eliminated with the inclusion of exchange (Fock) terms, which leads to a new CDF model --- the density-dependent relativistic Hartree-Fock (DDRHF) theory \cite{Long2006PLB1} and its natural extension, the relativistic Hartree-Fock-Bogoliubov (RHFB) theory \cite{Long2010PRC}. Besides compatible quantitative accuracy as RMF in describing nuclear bulk properties, substantial improvements are also achieved by DDRHF in the self-consistent description of nuclear shell structures and the evolutions \cite{Long2007PRC, Long2008EPL, Long2009PLB}, the relativistic symmetry restoration \cite{Long2006PLB2, Liang2010EPJA, Long2010PRCR}, low-energy excitation modes \cite{Liang2008PRL}, and neutron star physics \cite{Sun2008PRC, Long2012PRC} etc.

In general the pairing effects in open-shell nuclei are considered within the BCS or Bogoliubov schemes. For a stable nuclear system, the BCS method can provide an efficient and simple way to handle the pairing correlations while it meets serious problems going beyond the stable region, especially when the halos emerge \cite{Meng2006PPNP}. Approaching the nuclear isospin limits, the single neutron or proton separation energy becomes comparable to the pairing-gap energy such that the continuum can be got involved easily by the pairing correlations. In terms of Bogoiubov quasi-particles, both mean field and pairing correlations are unified into the Bogoliubov scheme such that the continuum effects can be naturally taken into account. Aiming at the weakly bound nuclear systems, the CDF models combined with Bogoliubov framework have been extended as the relativistic Hartree-Bogoliubov (RHB) theory \cite{Meng1998NPA, Vretenar2005PhysRep} and the RHFB theory \cite{Long2010PRC}. Compared to the former, the RHFB theory provides a more concrete platform to evaluate the pairing effects due to the improvements brought by the Fock terms, especially with the inclusion of $\rho$-tensor couplings \cite{Long2007PRC}.

Besides the self-consistent treatment of the continuum effects, e.g., by the Bogoliubov theory, the theoretical reliability in describing the pairing effects also depends on the adopted pairing interaction. In the realistic calculations the particle-particle (p-p) pairing interactions are usually taken as a phenomenological form, such as zero-range $\delta$-forces or finite-range Gogny interactions with great success in the relativistic and non-relativistic calculations \cite{Skyrme1956PM, Berger1991CPC, Meng2006PPNP, Long2010PRC}. Due to its numerical simplicity, the $\delta$-type forces are popularly applied in dealing with the pairing effects, especially when a realistic density dependence is introduced \cite{Dobaczewski1996PRC}. While limited by the zero-range formalism, it may require a sophisticated energy cut-off to decide the pairing window, as well as the strength of pairing interaction \cite{Dobaczewski1996PRC}. In contrast to the simple $\delta$-force, the finite-range Gogny forces have been widely taken as the effective interactions not only in the p-p channels but also in the particle-hole (p-h) channels and potentially better systematics is expected for the p-p interaction. In addition benefited from the characteristic finite range, a natural energy cut-off is already embedded in modeling the pairing space. While it should be noticed that the pairing parts of Gogny forces were adjusted in consistent with the non-relativistic mean fields \cite{Berger1991CPC, Chappert2008PLB, Goriely2009PRL}. The global consistence associated with the relativistic scheme \cite{Gonzalez1996PLB, Afanasjev2003PRC} is still required to be tested, especially when the Fock terms are included in the p-h channels.

In this work, the OESs on nuclear binding energy will be systematically studied under the CDF scheme, specifically the RHFB and RHB theories. The content is organized as follows. In Sec. \ref{rhfb}, we briefly introduce the general formalism of the RHFB theory with finite-range Gogny pairing force. The neutron OESs along the isotopic chains of C, O, Ca, Ni, Zr, Sn, Ce, Gd and Pb, and the proton ones along the isotonic chains of $N=50$, $82$ will be discussed systematically in Sec. \ref{OES resul}, as well as the effects with modified pairing strength. Finally, the summary is given in Sec. \ref{sum}.

\section{Theoretical formalism and numerical details}\label{rhfb}

Relativistically the nucleon-nucleon interaction in nuclear system is mediated by the exchange of mesons and photons. Standing on this criterion, the CDF model Lagrangian, i.e., the theoretical starting point, is constructed by including the degrees of freedom associated with the nucleon ($\psi$), the isoscalar $\sigma$- and $\omega$-mesons, the isovector $\rho$- and $\pi$-mesons, and the photons ($A$) \cite{Bouyssy1987PRC}. Following the standard variational procedure, the system Hamiltonian $H$ is then determined as well as the field equations for nucleons, mesons and photons, respectively the Dirac, Klein-Gordon and Proca equations \cite{Serot:1986, Bouyssy1987PRC}.

Standing on the level of mean field approach, the contributions of the negative energy states are generally neglected, namely the no-sea approximation. The CDF energy functional is then determined by the expectation of the system Hamiltonian $H$ with respect to the Hartree-Fock ground $\lrlc{\Phi_0}$,
\begin{align}\label{Energy Fucntional}
E =& \langle \Phi_0| {H}|\Phi_0\rangle,&\lrlc{\Phi_0} = & \prod_{i=1} c_i^\dag\lrlc{0}
\end{align}
where the index $i$ denotes the states with positive energy and $\lrlc{0}$ is the vacuum state. Within the RHB theory, the energy functional (\ref{Energy Fucntional}) contains only the Hartree contributions, whereas both Hartree and Fock terms are included explicitly in the RHFB theory \cite{Meng1998NPA, Long2010PRC}.

In the spherical system, the variation of the energy functional (\ref{Energy Fucntional}) with respect to the Dirac spinor $\psi(\bm{r})$ leads to the relativistic Hartree-Fock (RHF) equation as
\begin{align}\label{DHF}
\int d\bm{r}^\prime h(\bm{r},\bm{r}^\prime) \psi(\bm{r}^\prime) =& \epsilon\psi(\bm{r}),
\end{align}
where $\epsilon$ is the single-particle energy and $h(\bm{r},\bm{r}^\prime)$ denotes the single-particle Hamiltonian \cite{Long2010PRC}. Combining the Bogoliubov transformation \cite{Gorkov1958} with CDF models, the above RHF equation can be extended into the RHFB one \cite{Kucharek1991ZPA} as
\begin{equation}
\label{RHFB}
\begin{split}
\int d\bm{r}' & \left( \begin{array}{cc} h(\bm{r},\bm{r}^\prime) & \Delta(\bm{r},\bm{r}^\prime) \\[0.5em]
- \Delta(\bm{r},\bm{r}^\prime) & h(\bm{r},\bm{r}^\prime) \end{array} \right)
\left(\begin{array}{c} \psi_U(\bm{r}') \\[0.5em] \psi_V(\bm{r}')\end{array}\right) \\
&\hspace{4em} = \lrb{\begin{matrix}\lambda + E_q &0\\[0.5em] 0&\lambda-E_q\end{matrix}}\left(\begin{array}{c} \psi_U(\bm{r})\\[0.5em]\psi_V(\bm{r})\end{array}\right),
\end{split}
\end{equation}
where $\psi_U$ and $\psi_V$ are the quasi-particle spinors, $E_q$ denotes the quasi-particle energy, and the chemical potential $\lambda$ is introduced to preserve the particle number on the average. In the single-particle Hamiltonian $h(\bm{r},\bm{r}^\prime)$, the retardation effects are neglected as is usually done in mean-field calculations \cite{Long2010PRC}. The pairing potential $\Delta(\svec r, \svec r')$ in Eq. (\ref{RHFB}) reads as
\begin{align}\label{pairing potential}
\Delta_\alpha(\bm{r},\bm{r}^\prime)=&-\frac{1}{2} \sum_\beta V^{pp}_{\alpha\beta}(\bm{r},\bm{r}^\prime)\kappa_\beta (\bm{r},\bm{r}^\prime),
\end{align}
with the pairing tensor
\begin{align}\label{pairing tensor}
\kappa_\alpha(\bm{r},\bm{r}^\prime) =& \psi_{V_\alpha}(\bm{r})^\ast \psi_{U_\alpha}(\bm{r}^\prime).
\end{align}
For the p-p interaction $V^{pp}$ in Eq. (\ref{pairing potential}), the following finite-range Gogny force is adopted in this work,
\begin{equation}\label{Gogny}
%\begin{split}
V(\bm{r},\bm{r}') =\sum_{i=1,2} e^{((r-r^\prime)/\mu_i)^2}(W_i +B_iP^\sigma -H_iP^\tau -M_iP^\sigma P^\tau) ,
%\end{split}
\end{equation}
where $\mu_i, W_i, B_i, H_i$, and $M_i$ are the parameters of Gogny pairing force.

Except for few special cases, e.g., the RHB model with zero-range pairing force, the RHFB equation (\ref{RHFB}) is generally in an integro-differential form, which is difficult to be solved efficiently in the coordinate space. In this work, the Dirac Woods-Saxon (DWS) basis \cite{Zhou2003PRC}, which can provide appropriate asymptotical behavior for the continuum states, is introduced to solve such integro-differential equation. For further details, the readers are referred to Ref. \cite{Long2010PRC}.

To evaluate the pairing effects, the following three-point indicators \cite{Bender2000EPJA, Dobaczewski2001PRC} of the OES of nuclear binding energies are introduced for isotopes ($\nu$) and isotones ($\pi$), respectively,
\begin{subequations}\label{three point}
\begin{align}
\Delta^{(3)}_\nu (N) \equiv& \frac{(-1)^N}{2}\lrs{S_n(N,Z)-S_n(N+1,Z) }, \\
\Delta^{(3)}_\pi (Z) \equiv& \frac{(-1)^Z}{2}\lrs{S_p(N,Z)-S_p(N,Z+1) },
\end{align}
\end{subequations}
where $\Delta^{(3)}_\nu (N)$ and $\Delta^{(3)}_\pi (Z)$ represent neutron ($\nu$) and proton ($\pi$) OES, i.e., the difference on single-nucleon separation energies $S_n$ (or $S_p$) of two neighboring isotopes (or isotones). Quantitatively there also exist other higher order indicators to evaluate the pairing effects, such as the four-point formula, or the five-point ones \cite{Bender2000EPJA}
\begin{subequations}\label{five point}
\begin{align}
\Delta^{(5)}_\nu (N) \equiv& \frac{(-1)^{N}}{8}[S_n(N+2,Z)-3S_n(N+1, Z) \nonumber\\
& \qquad\qquad  +3S_n(N,Z)-S_n(N-1,Z)], \\
\Delta^{(5)}_\pi (Z) \equiv& \frac{(-1)^{Z}}{8}[S_p(N,Z+2)-3S_p(N, Z+1) \nonumber\\
& \qquad\qquad  +3S_p(N,Z)-S_p(N,Z-1)],
\end{align}
\end{subequations}
However, these higher order indicators may get the mean-field features involved, which may partially smooth out the odd-even oscillations due to the pairing effects \cite{Dobaczewski2001PRC}. Even with the three-point one (\ref{three point}), the obtained OESs for the isotopes (isotones) with even neutron (proton) number are still sensitive to both pairing and mean-field effects \cite{Satula1998PRL, Dobaczewski2001PRC, Bertsch2009PRC}. In order to avoid the disturbances of the mean field as much as possible, here we only take the OESs [see Eq. (\ref{three point})] of the isotopes (isotones) with odd neutron (proton) number as the observables to study the pairing effects.

To have an overall understanding on the pairing properties, we performed the calculations within both RHFB and RHB for the isotopes of C, O, Ca, Ni, Zr, Sn, Ce, Gd and Pb, and the isotones of $N=50$ and $82$, from which are extracted the neutron and proton OESs by Eq. (\ref{three point}) for the odd isotopes and isotones, respectively. In all the calculations, the p-p interaction $V^{pp}$ is taken as the finite-range Gogny force D1S \cite{Berger1984NPA}. In the mean field channels, we adopt four CDF effective interactions, specifically PKA1 \cite{Long2007PRC} and PKO3 \cite{Long2008EPL} within RHFB, and PKDD \cite{Long2004PRC} and DD-ME2 \cite{Lalazissis2005PRC} within RHB. For the selected isotopes and isotones, it is checked to be accurate enough to take the parameters of the DWS basis as $R_{\text{max}}=20$ fm, $N_F=28$ and $N_D=12$, where $R_{\text{max}}$ is the size of the spherical box and $N_F(N_D)$ corresponds to the numbers of positive(negative) energy states included in the basis expansion. For the nuclei with odd neutron and/or proton numbers, the blocking configurations are utilized here to approximate the odd-nucleon effects. Practically by blocking different orbits around the Fermi surface the configuration with the strongest binding is determined as the ground state. Notice that in the odd nuclei the time reverse symmetry is broken due to the odd nucleon. While the time-odd mean fields are poorly known and their effects differ from model to model. In this work we neglect the current effects since the mean field would not be essentially changed by the odd particle \cite{Afanasjev2010PRC}.

\section{\label{OES resul}RESULTS AND DISCUSSIONS}

\subsection{Odd-even staggering along the isotopic and isotonic chains}

\begin{figure*}[htbp]
\ifpdf
\includegraphics[width=0.9\textwidth]{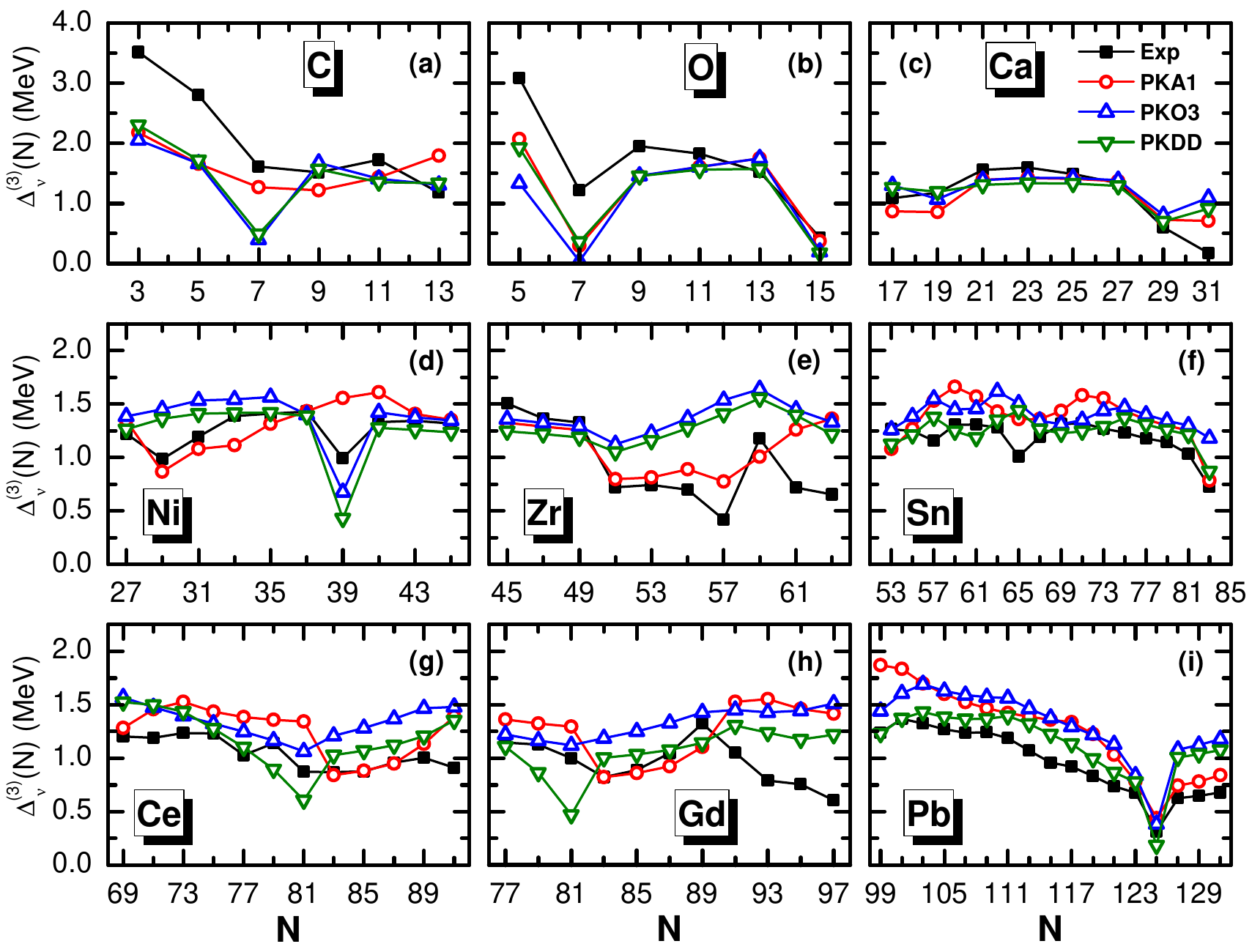}
\else
\includegraphics[width=0.9\textwidth]{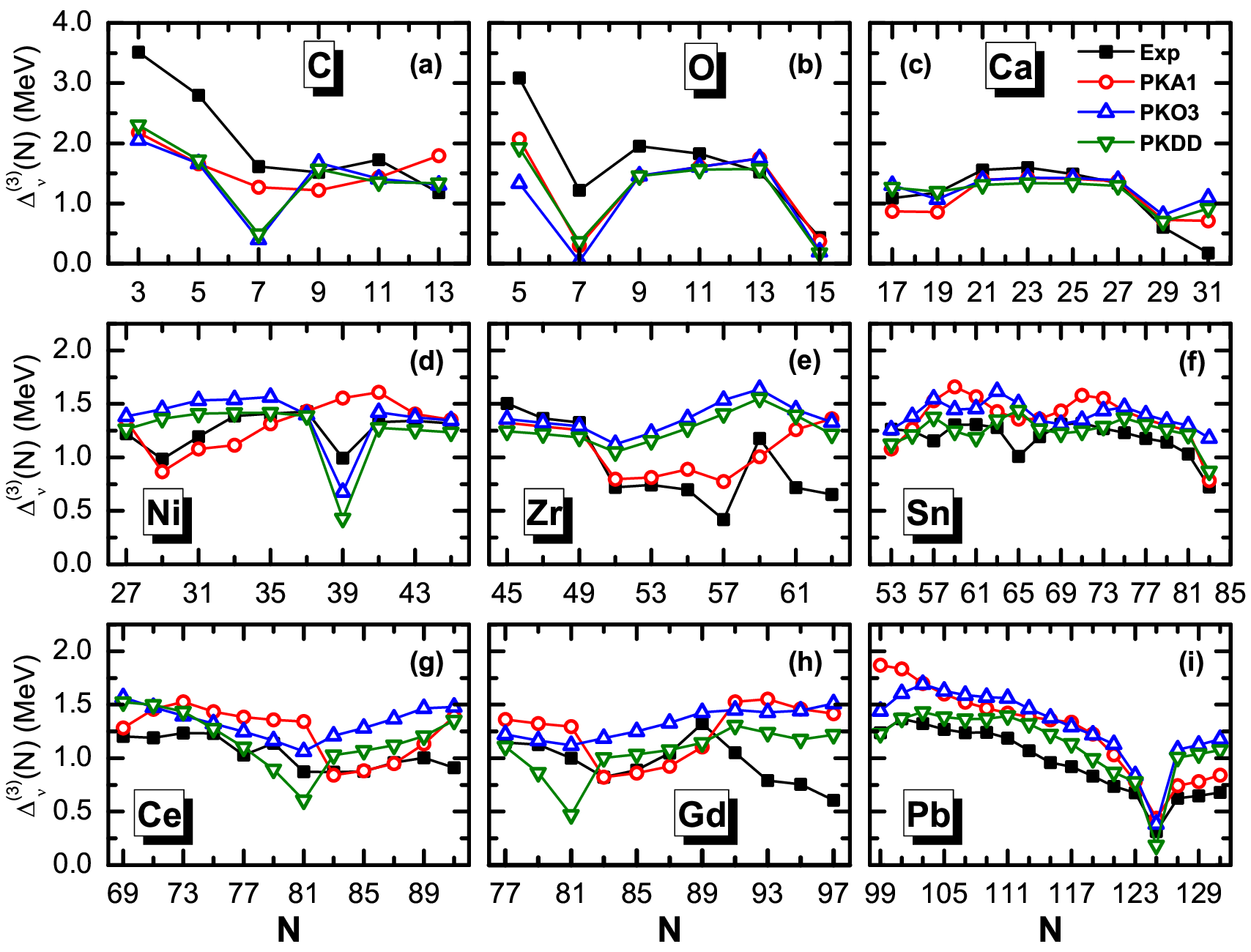}
\fi
\caption{\label{fig:OES} (Color online) Neutron OESs $\Delta^{(3)}_\nu(N)$ [see Eq. (\ref{three point})] as functions of neutron number $N$ for the C, O, Ca, Ni, Zr, Sn, Ce, Gd, and Pb isotopes. The results are extracted from the calculations of RHFB with PKA1 (in open circles) and PKO3 (in open up-triangles), and RHB with PKDD (in open down-triangles), in comparison with the data \cite{Audi2003NPA} (in filled squares). The finite-range Gogny force D1S is utilized as the effective pairing force. See the text for details.}
\end{figure*}

We first study the neutron OESs of the selected isotopes with odd neutron number within the RHFB and RHB theories. Figure \ref{fig:OES} shows the evolution of neutron OESs $\Delta^{(3)}_\nu(N)$ (in open symbols) along the isotopic chains of C, O, Ca, Ni, Zr, Sn, Ce, Gd and Pb, where the experimental data (in filled squares) extracted from Ref. \cite{Audi2003NPA} are presented as the reference. The theoretical results are provided by the calculations with PKA1, PKO3 and PKDD. The results from DD-ME2 are omitted because they show similar systematics as PKDD on the OESs. It is clearly seen that all the calculations with the selected effective interactions can provide appropriate overall agreement with the data to certain extent.

For the light nuclei, PKA1 shows better agreements with the data than PKO3 and PKDD on the isospin evolution of OESs along the isotopic chain of C. For O isotopes, the experimental depressions of neutron OESs $\Delta^{(3)}_\nu(N)$ at $N=7$ and $N=15$ are well reproduced by all the theoretical calculations, consistent with the occurrences of the shell structures at $N=8$ and $16$. While all the theoretical calculations systematically underestimate the OESs $\Delta^{(3)}_\nu(N)$ for both C and O isotopes, especially at the proton rich sides.

For the medium-heavy nuclei it is found that the neutron OESs of Ca isotopes are precisely reproduced by PKA1, PKO3 and PKDD only except the one at $N=31$. For Ni isotopes, PKA1, PKO3 and PKDD also present proper overall agreements, except $N=29$ for PKO3 and PKDD, and $N=39$ for all the selected effective interactions. From the consistent relation between the pairing effects and shell structures, it seems that the shell effect around $N=28$ is underestimated by PKO3 and PKDD, whereas the one at $N=40$ is underestimated by PKA1 and overestimated by PKO3 and PKDD. For Zr isotopes, the neutron OESs are described precisely by PKA1 until the distinct discrepancies occur beyond $N=59$, where the nuclei may be deformed \cite{Moller1995ADNDT}. In contrast PKO3 and PKDD present remarkable deviations from the data in a fairly wide range since $N>49$. The model deviations in reproducing the neutron OESs here may be connected with the fact that PKA1, in which the $\rho$-tensor coupling is included \cite{Long2007PRC}, provides better descriptions on the nuclear shell structures and therefore improves the consistence between the pairing and mean-field effects, as mentioned above.

For heavier nuclei, e.g., the Sn isotopes, the experimental OESs $\Delta^{(3)}_\nu(N)$ keep around the value of 1.2 MeV, except at $N=65$ and $N=83$. Quantitatively PKDD provides the best agreement with the data while the neutron OESs are slightly overestimated by PKA1 and PKO3. For Ce and Gd isotopes, although many of them may be deformed \cite{Moller1995ADNDT}, the spherical calculations with PKA1, PKO3 and PKDD still provide appropriate descriptions on the neutron OESs, except for Gd isotopes beyond $N=89$. Along the isotopic chain of Pb, the systematics of neutron OESs are properly described by PKA1, PKO3 and PKDD, while there still exists some systematical overestimation.

\begin{figure}[htbp]
\ifpdf
\includegraphics[width=0.48\textwidth]{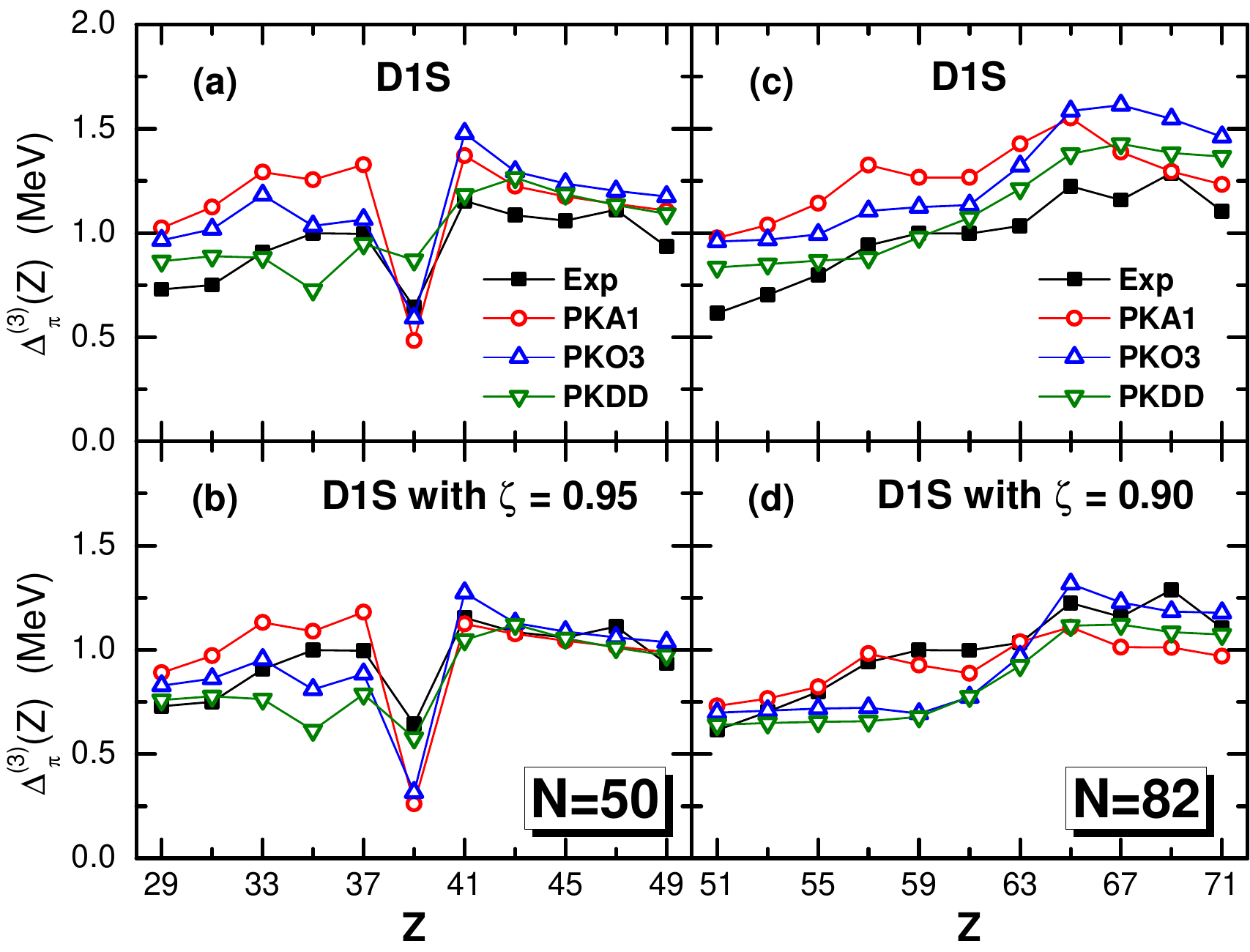}
\else
\includegraphics[width=0.48\textwidth]{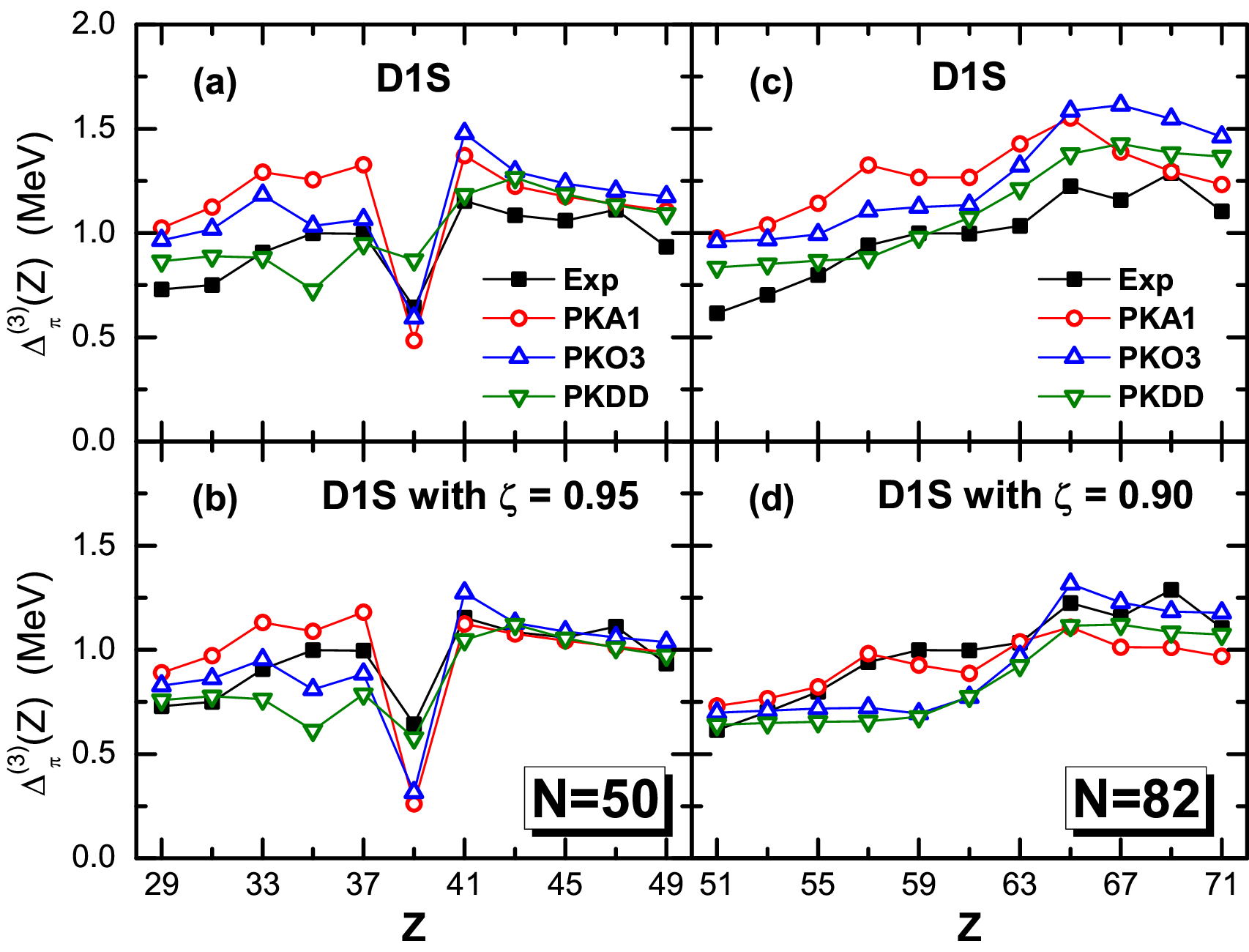}
\fi
\caption{\label{fig:isotones} (Color online) Proton OESs $\Delta^{(3)}_\pi(Z)$ of $N=50$ and $82$ isotones as functions of proton number $Z$. The results are calculated by the RHFB theory with PKA1 and PKO3, and by the RHB theory with PKDD, where the Gogny force D1S is utilized in the pairing channel. The experimental data extracted from Ref. \cite{Audi2003NPA} are also shown for comparison. The results without and with the effective pairing strength factor $\zeta$ are shown in the upper ((a) and (c)) and lower ((b) and (d)) panels respectively.}
\end{figure}

Concerning the evolutions of proton OESs along the isotonic chains of $N=50$ and $82$, similar systematics are also obtained from the CDF calculations with PKA1, PKO3 and PKDD, as referred to the data \cite{Audi2003NPA}. From Fig. \ref{fig:isotones} (a) and (c) it is seen that the proton OESs for $N=50$ and $82$ isotones are reproduced in an appropriate agreement with data by the selected effective interactions. In particular, the calculations with both PKA1 and PKO3 present the clear depression of $\Delta^{(3)}_\pi(Z)$ at $Z=39$ for $N=50$ isotones, corresponding with the sub-shell structure $Z=40$, which may imply the extra role of Fock terms especially the embedded tensor effects in determining the nuclear shell structure for the selected isotonic chain \cite{Wang2012tensor}. Compared to the data quantitatively, however, the pairing effects are also somewhat overestimated for the selected isotones. It is worthwhile to mention that the calculations with other relativistic effective Lagrangian, e.g., the RHB with DD-ME2 \cite{Lalazissis2005PRC}, also show similar trend as above. In brief one may reach the point that the OESs, coherently the pairing effects, are somewhat underestimated by the CDF calculations for light nuclei and overestimated for heavy nuclei.

\begin{table*}[htbp]
\caption{The root-mean-square deviation $\sigma_{\textrm{rms}}$ and the average error $\mathcal {D}$ of OES for the isotopes C, O, Ca, Ni, Zr, Sn, Ce, Gd, Pb, and for the isotones $N=50$ and $82$, determined by the CDF calculations with PKA1, PKO3 and PKDD. The results inside/outside the parenthesis are presented by the calculations taking the Gogny force D1S with/without pairing strength factor $\zeta$ as the effective pairing interaction. See the text for details.}\label{tab:factors}
\begin{ruledtabular}
\begin{tabular}{cccccccc}
     & \multirow{2}{*}{$\zeta$} & \multicolumn{3}{c}{$\sigma_{\textrm{rms}}$} & \multicolumn{3}{c}{$\mathcal {D}$}  \\ \cline{3-5}\cline{6-8}
     &         &      PKA1      &      PKO3      &      PKDD      &       PKA1       &       PKO3       &       PKDD        \\ \hline
C    & $1.20$  & $0.792(0.665)$ & $0.916(0.816)$ & $0.824(0.671)$ & $-0.467(-0.125)$ & $-0.640(-0.203)$ & $-0.597(-0.159)$  \\
O    & $1.15$  & $0.609(0.582)$ & $0.899(0.649)$ & $0.641(0.594)$ & $-0.415(-0.173)$ & $-0.609(-0.100)$ & $-0.498(-0.173)$  \\
Ca   & $1.00$  & $0.255       $ & $0.354       $ & $0.305       $ & $-0.032        $ & $+0.109        $ & $+0.039        $  \\
Ni   & $1.00$  & $0.230       $ & $0.228       $ & $0.230       $ & $+0.046        $ & $+0.108        $ & $-0.014        $  \\
Zr   & $1.00$  & $0.323       $ & $0.577       $ & $0.511       $ & $+0.128        $ & $+0.393        $ & $+0.295        $  \\
Sn   & $0.95$  & $0.230(0.135)$ & $0.257(0.145)$ & $0.158(0.167)$ & $+0.186(+0.043)$ & $+0.211(-0.085)$ & $+0.062(+0.035)$  \\
Ce   & $0.93$  & $0.266(0.155)$ & $0.332(0.176)$ & $0.241(0.184)$ & $+0.205(-0.002)$ & $+0.294(+0.083)$ & $+0.132(-0.037)$  \\
Gd   & $0.90$  & $0.447(0.339)$ & $0.452(0.251)$ & $0.338(0.314)$ & $+0.281(-0.002)$ & $+0.362(+0.056)$ & $+0.098(-0.160)$  \\
Pb   & $0.90$  & $0.327(0.118)$ & $0.362(0.122)$ & $0.218(0.163)$ & $+0.296(+0.049)$ & $+0.344(+0.057)$ & $+0.167(-0.088)$  \\ %\hline
N=50 & $0.95$  & $0.250(0.172)$ & $0.204(0.137)$ & $0.148(0.148)$ & $+0.196(+0.037)$ & $+0.171(-0.011)$ & $+0.062(-0.079)$  \\
N=82 & $0.90$  & $0.300(0.123)$ & $0.287(0.147)$ & $0.163(0.171)$ & $+0.278(-0.055)$ & $+0.269(-0.061)$ & $+0.127(-0.135)$  \\
\end{tabular}
\end{ruledtabular}
\end{table*}

In general the statistical observables, i.e., the root-mean-square (rms) deviation $\sigma_{\text{rms}}$ and the average one $\mathcal D$ from the data, are introduced to test quantitative precision of the theoretical calculations. Table~\ref{tab:factors} shows the values of $\sigma_{\textrm{rms}}$ and $\mathcal {D}$ for the selected isotopes and isotones. It is found that the calculations with PKA1, PKO3 and PKDD present negative average error $\mathcal {D}$ for C and O isotopes and positive ones for nuclei heavier than Ni isotopes. Such systematics may also indicate that the pairing effects are somehow underestimated for light nuclei and overestimated for heavy nuclei by taking the finite-range Gogny force D1S as the effective pairing interaction, which has already been demonstrated from the results in Fig. \ref{fig:OES} and Fig. \ref{fig:isotones} (a, c).

\begin{figure}[htbp]
\ifpdf
\includegraphics[width=0.48\textwidth]{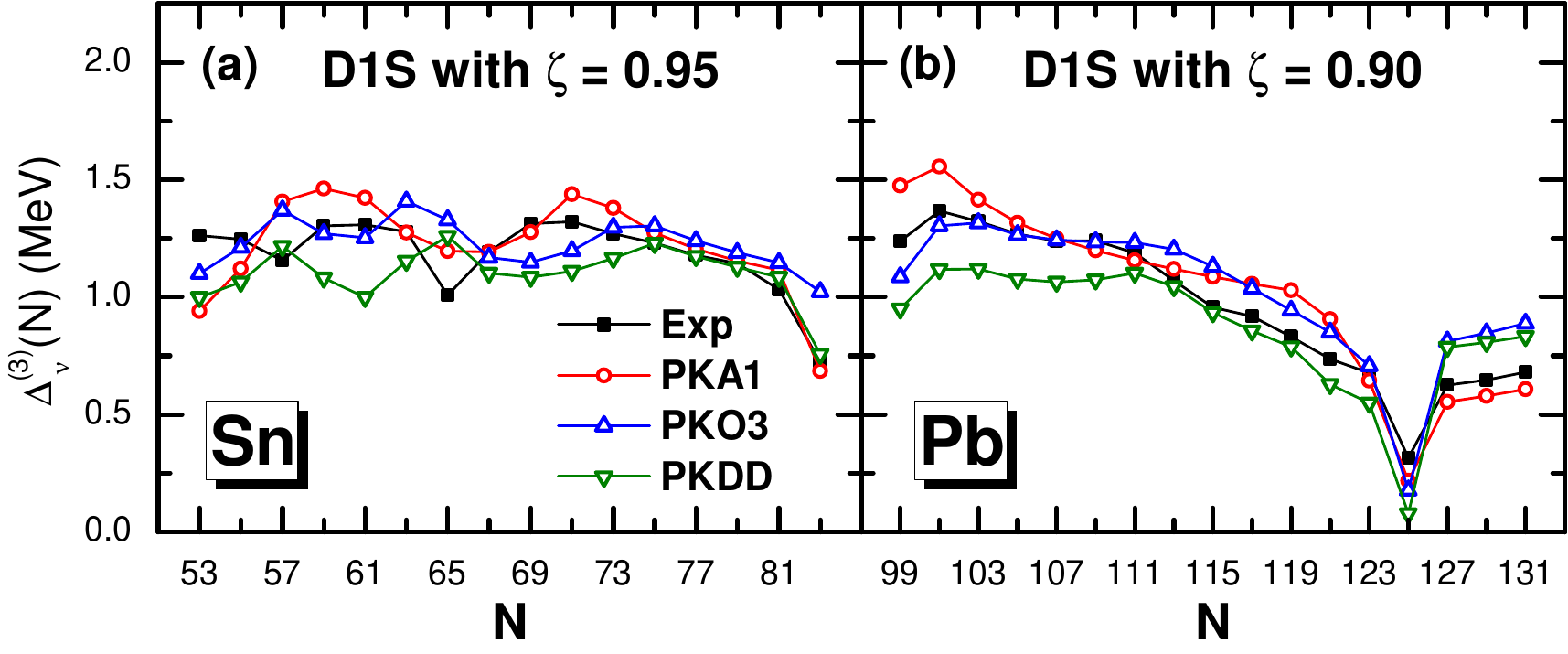}
\else
\includegraphics[width=0.48\textwidth]{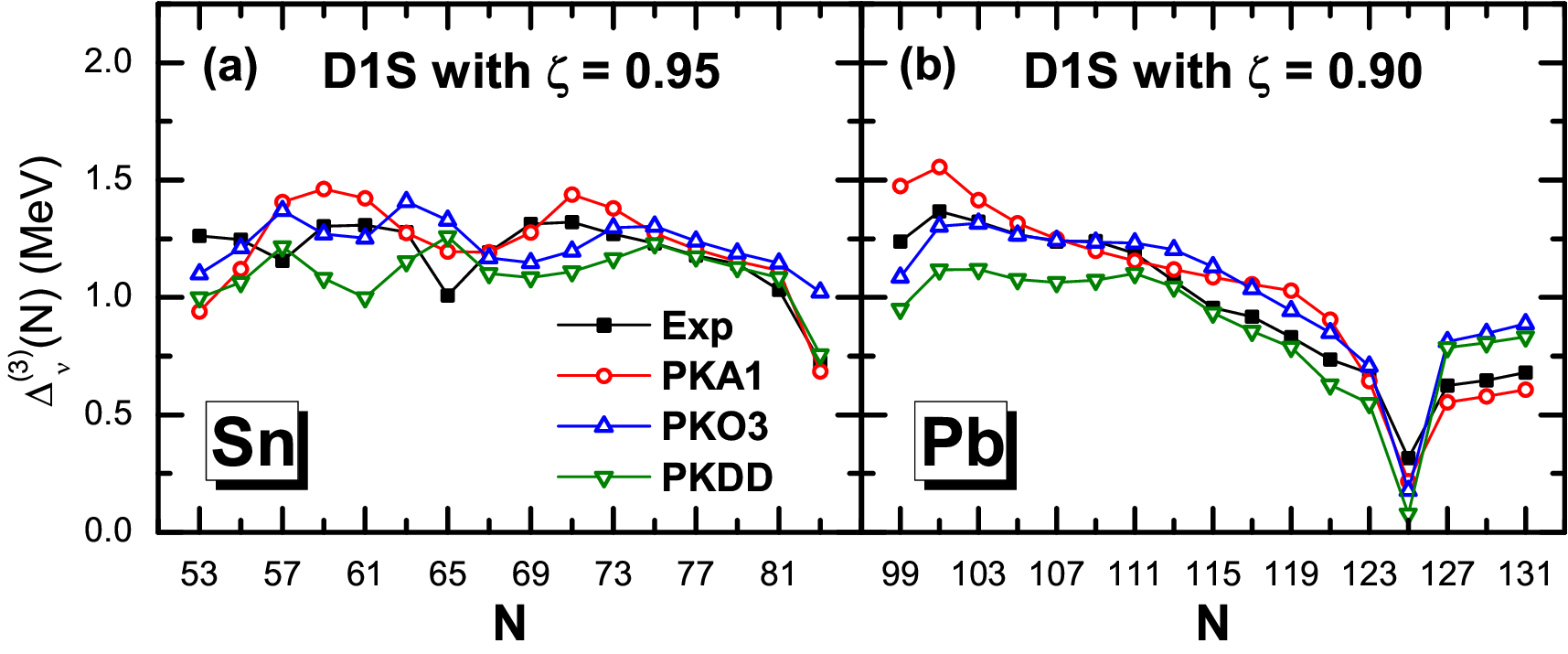}
\fi
\caption{\label{fig:SnPb} (Color online) The same as Fig. \ref{fig:OES} but for Sn (panel (a)) and Pb (panel(b)) isotopes calculated by the optimized Gogny pairing force D1S with the effective pairing factor $\zeta$. See the text for details.}
\end{figure}

To eliminate such systematical discrepancy with the data, here we consider a simple modification by introducing an additional strength factor $\zeta$ into the pairing interaction, similar as in Ref. \cite{Gonzalez1996PLB},
\begin{eqnarray}\label{zeta}
V_{\text{opt}}^{\text{pp}}(\bm{r},\bm{r}^\prime) = \zeta V^{\text{pp}}(\bm{r},\bm{r}^\prime),
\end{eqnarray}
where $V^{\text{pp}}(\bm r, \bm r')$ and $V_{\text{opt}}^{\text{pp}}(\bm r, \bm r')$ are the original and modified pairing force. The strength factor $\zeta$ is optimized with respect to the root-mean-square deviation $\sigma_{\textrm{rms}}$ from the PKA1 calculations with different values of $\zeta$ since PKA1 can better describes the nuclear structure properties \cite{Long2007PRC}. As the examples, Fig. \ref{fig:SnPb} shows the neutron OESs of Sn and Pb isotopes calculated with the optimized Gogny pairing force D1S, in which the strength factor $\zeta$ are respectively taken as 0.95 and 0.90. Comparing to the results with original pairing force (i.e., $\zeta=1.0$) in Fig. \ref{fig:OES} (f) and (i), one can see that the agreements with the data are remarkably improved with the inclusion of the strength factor, where the overestimation on the neutron OES is reduced remarkably. In Ref. \cite{Afanasjev2003PRC} similar conclusions have also been reached in reproducing the experimental data of inertia moments for very heavy nuclei. For the $N=50$ and $82$ isotones, as shown in Fig. \ref{fig:isotones} (b) and (d), the systematical overestimation on proton OESs can be also eliminated by reducing the strength factors as $\zeta = 0.95$ and $0.90$, respectively.

In fact, such improvements can be demonstrated quantitatively from the statistical observables in Table \ref{tab:factors} for all the selected isotopes and isotones. It is shown that with the inclusion of strength factor almost all the CDF calculations present nearly zero average deviations $\mathcal D$, especially for the heavy isotopes and isotones, which corresponds to a systematic improvement on the description of the pairing effects.  As shown in the second column of Table \ref{tab:factors}, the strength factors $\zeta$ represent distinct $Z$-dependence for isotopes, as well as $N$-dependence for isotones, i.e., $\zeta$ decreases monotonically with increasing proton number $Z$ of the isotopes or neutron number $N$ of the isotones, consistent with the systematic deviations as seen from Fig. \ref{fig:OES} and \ref{fig:isotones}. As seen from the rms deviations in Table \ref{tab:factors}, remarkable improvements on the agreements can be also found with the inclusion of the strength factor in p-p interaction. For the neutron OESs the statistic qualities of the agreements, i.e., the values of $\sigma_{\textrm{rms}}$, are improved by about $10\%$ for light nuclei (C and O) and about $30\%$ for heavy ones (Sn, Ce, Gd and Pb). Whereas for the isotones of $N=50$ and $N=80$ (the last two rows in Table \ref{tab:factors}), the RHFB calculations with optimized pairing forces respectively present about $30\%$ and $50\%$ improvement on the rms deviations $\sigma_{\textrm{rms}}$, which keep almost unchanged in the RHB calculations with PKDD as well as with DD-ME2.

%\begin{figure}[htbp]
%\ifpdf
%\includegraphics[width=0.48\textwidth]{fig4.pdf}
%\else
%\includegraphics[width=0.48\textwidth]{fig4.eps}
%\fi
%\caption{\label{fig:fig4} (Color online) The average neutron pairing gap of Sn and Pb even isotopes compared with the corresponding neutron OES extracted from the data \cite{Audi2003NPA}. The results without and with the effective pairing strength factor $\zeta$ are shown in the left ((a) and (c)) and right ((b) and (d)) panels respectively.}
%\end{figure}

\begin{figure}[htbp]
\ifpdf
\includegraphics[width=0.48\textwidth]{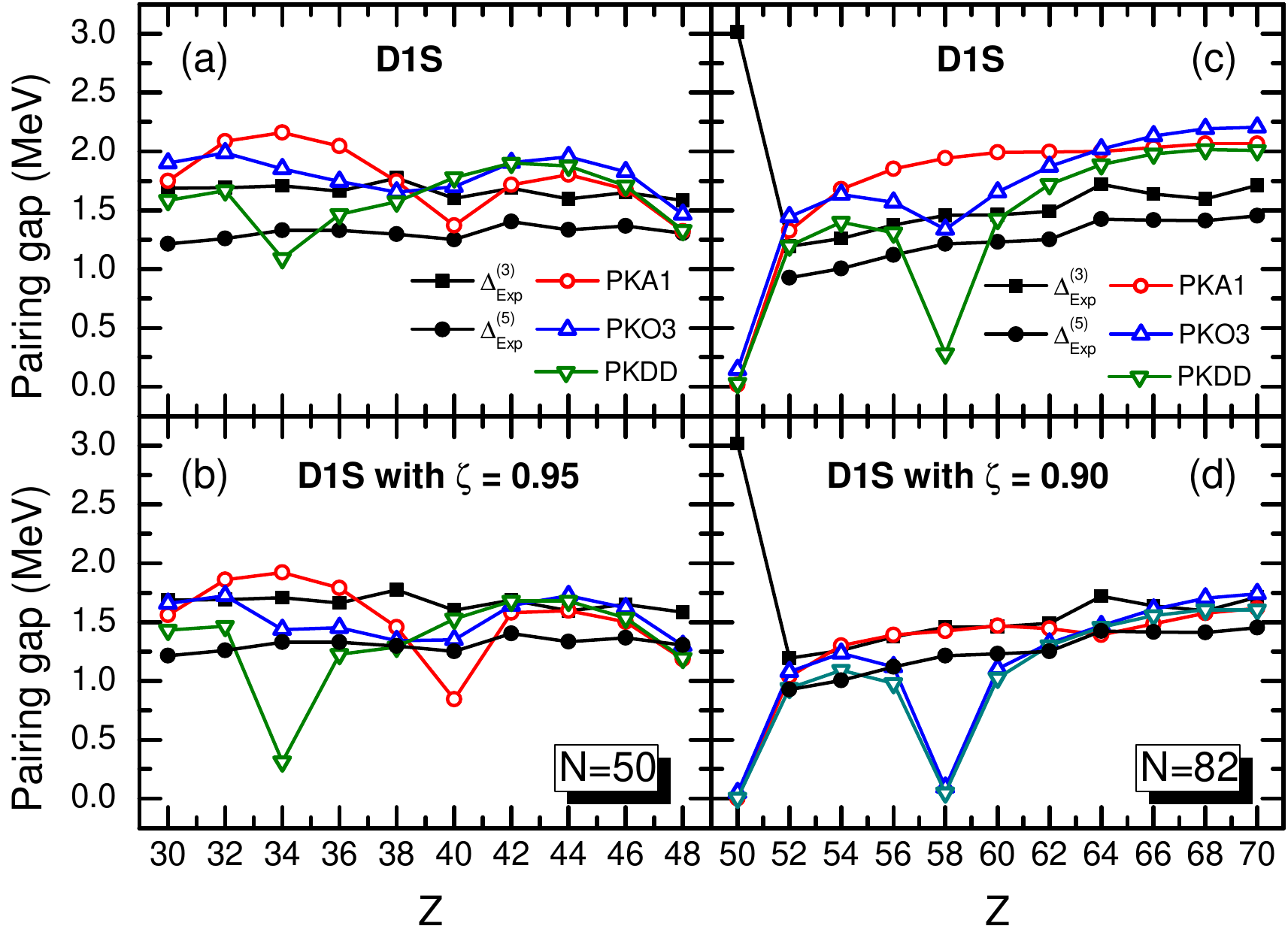}
\else
\includegraphics[width=0.48\textwidth]{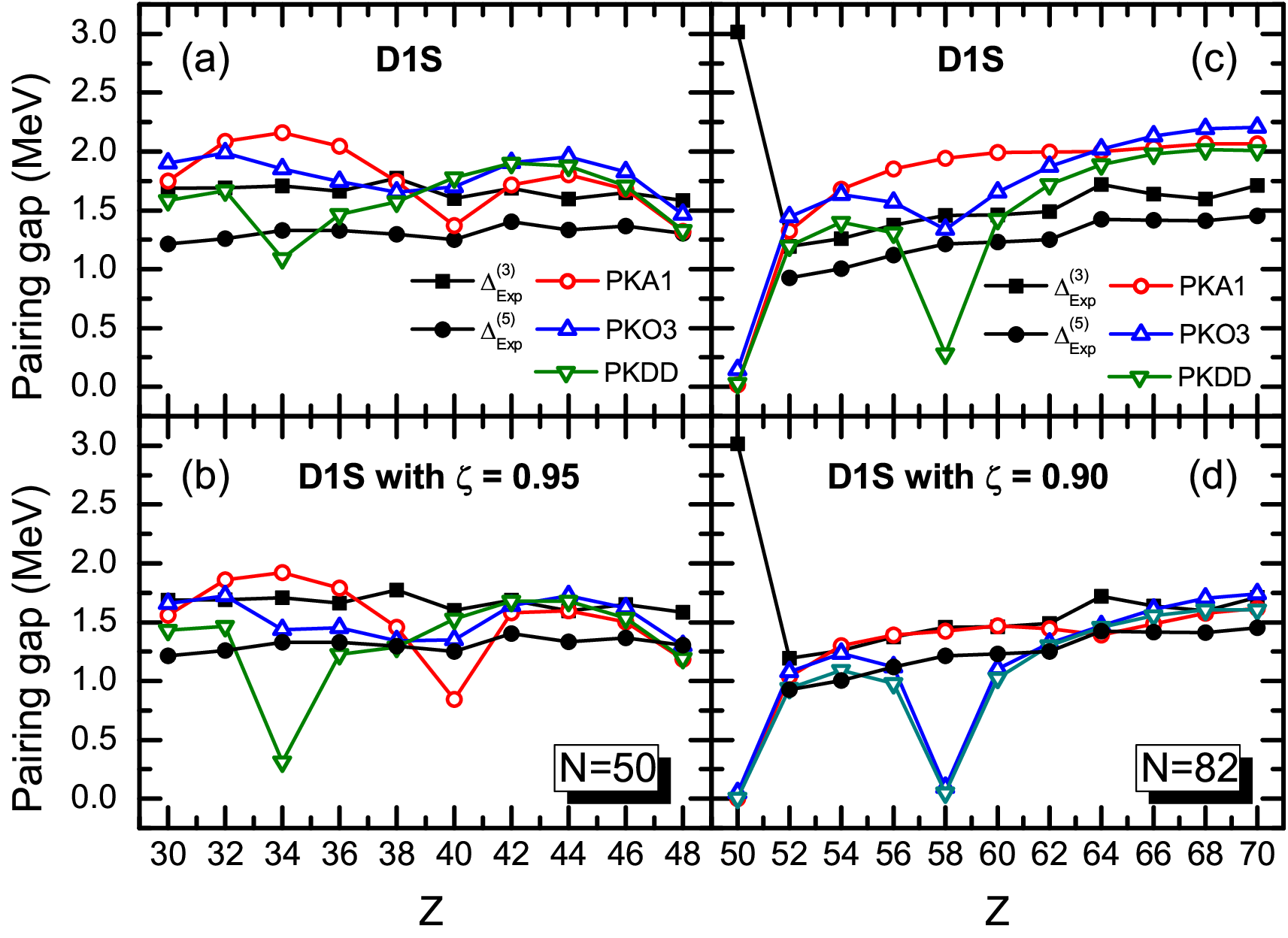}
\fi
\caption{\label{fig:fig5} (Color online) The average proton pairing gap of $N=50$ and $N=82$ even isotones compared with the corresponding neutron OES extracted from the data \cite{Audi2003NPA}. The results without and with the effective pairing strength factor $\zeta$ are shown in the left ((a) and (c)) and right ((b) and (d)) panels respectively. See the text for details. }
\end{figure}

As an implement, Fig. \ref{fig:fig5} presents the proton effective pairing gaps, i.e., the average gap energies extracted from the RHFB and RHB calculations, for the $N=50$ (left plots) and $82$ (right plots) isotones with even proton numbers. For comparison, are also shown the OESs of the experimental binding energies by three-point (filled squares) and five-point (filled circles) indicators. As the experimental reference of pairing effects, one may find distinct deviations between these two indicators along the isotonic chain of $N=50$ [see Fig. \ref{fig:fig5} (a)] and such deviations become smaller for $N=82$ isotones. This is mainly due to the fact that these evaluations of pairing effects for even-even nuclei have already got the mean-field effects involved to different extents \cite{Dobaczewski2001PRC, Satula1998PRL}. As referred to the experimental values, the calculated effective pairing gaps still show appropriate agreements, especially for the calculations with optimized pairing force. It should be noticed that along the isotonic chain of $N=50$ the calculations with PKA1 present distinct depression at $Z=40$, which implies the occurrence of some sub-shell closures, also demonstrated by the OES depression of the odd isotones at $Z=39$ [see Fig. \ref{fig:isotones} (a) and (b)]. While the calculations with PKO3 and PKDD do not show similar consistence between the results of odd and even isotones. E.g., for the odd $N=50$ isotones the OES results show certain depression at $Z=39$ [see Fig. \ref{fig:isotones} (a) and (b)] whereas the effective pairing gaps for the even isotones are just changed smoothly at $Z=40$ [see Fig. \ref{fig:fig5} (a) and (b)]. Along the isotonic chain of $N=82$, the inconsistence are also observed at $Z=58$ from the results of PKO3 and PKDD, i.e., the values of the effective pairing gaps clearly indicate the existence of the artificial shell closure $Z=58$ \cite{Geng2006CPL, Long2007PRC} while such spurious shell is not indicated by the systematics on the OESs of the odd isotones. From this aspect one may find that there exists some inconsistence between the relativistic effective Lagrangians (PKO3 and PKDD) and non-relativistic pairing treatment.

\subsection{Nuclear structure properties with optimized pairing force}

Since substantial improvements in reproducing the neutron/proton OESs have been achieved with the optimized pairing force $V_{\text{opt}}^{\text{pp}}$ [see Eq. (\ref{zeta})], it is worthwhile to test the relevant effects in describing the nuclear structure properties, such as the single-particle configurations, pairing energies and nuclear binding energies. As a direct response, the occupation probabilities of valence orbits will be essentially changed with the modifications of the pairing strength, as well as for the pairing energies.

In Table \ref{tab:Sn1} are shown the calculated occupation probabilities of neutron valence orbits $2d_{5/2}, 1g_{7/2}, 3s_{1/2}, 2d_{3/2}$ and $1h_{11/2}$ in $^{112}$Sn, $^{114}$Sn, $^{118}$Sn and $^{124}$Sn, as compared to the data taken from Refs. \cite{Andreozzi1996ZPA,Fleming}. The bracketed numbers are the results calculated with the optimized pairing force $V_{\text{opt}}^{\text{pp}}$ ($\zeta=0.95$). Among the selected effective Lagrangians, PKA1 shows the best agreements with the data within the error bars (denoted by bold types). Whereas less satisfied agreements are provided by the calculations with PKO3 and PKDD, especially the improper order of the pseudo-spin partners $2d_{5/2}$ and $1g_{7/2}$. As seem from the data of occupation probabilities, the neutron orbit $2d_{5/2}$ is almost fully filled whereas $1g_{7/2}$ is just gradually occupied, which implies that $2d_{5/2}$ is bound deeper than $1g_{7/2}$ in the selected isotopes. Evidently PKA1 presents more reliable description on nuclear structures than the other selected effective Lagrangians.

In the PKA1 calculations the optimized pairing force $V_{\text{opt}}^{\text{pp}}$ also bring some systematic improvements on the agreement with the data, i.e., being closer to the central values of the data. Whereas with PKO3 and PKDD, such systematical improvements can not be observed, which may be partially due to the inappropriate order of the valence orbits. If comparing the occupations of another pseudo-spin doublet $3s_{1/2}$ and $2d_{3/2}$, one may also find the improper order of these two states described by PKO3 and PKDD in $^{112, 118, 124}$Sn, and by DD-ME2 in $^{112}$Sn. Compared with the distinct improvements on the OESs with $V_{\text{opt}}^{\text{pp}}$, the corresponding effects on the occupations are relatively weak because the configurations are determined not only by the pairing effects, but also more essentially by the concrete shell structure, in which the mean field plays the dominant role.

\begin{table*}[htbp]
\caption{\label{tab:Sn1} Occupation numbers of valence neutron orbits in $^{112, 114,118,124}$Sn, calculated by RHFB with PKA1 and PKO3, and by RHB with PKDD and DD-ME2, in comparison with the data \cite{Andreozzi1996ZPA,Fleming}. In the calculations, the pairing force is adopted as the original Gogny D1S and the optimized one with $\zeta=0.95$ (inside the parenthesis), respectively. The bold types denote the theoretical results within the error bars of the data. See the text for details.}
\begin{ruledtabular}\setlength{\tabcolsep}{1pt}
\begin{tabular}{ccccccccccc}
\multicolumn{1}{c}{\multirow{2}{*}{Orbit}} & \multicolumn{5}{c}{$^{112}$Sn}  & \multicolumn{5}{c}{$^{114}$Sn}  \\ \cline{2-6} \cline{7-11}
       &      Exp.      &      PKA1      &      PKO3      &      PKDD      &DD-ME2&      Exp.     &      PKA1      &      PKO3      &      PKDD   &DD-ME2   \\ \hline
$\nu2d_{5/2}$  & $0.93\pm0.12$  &\emph{0.82}(\emph{0.84})&0.61(0.63)&0.61(0.62)&0.65(0.66)&$0.97\pm0.09$&0.86(\emph{0.88})&0.73(0.76)&0.759(0.80)&0.80(0.83)\\
$\nu1g_{7/2}$  & $0.63\pm0.13$  &\emph{0.55}(\emph{0.56})&0.83(0.86)&0.89(0.91)&0.86(0.88)&$0.69\pm0.15$&\emph{0.65}(\emph{0.67})&0.88(0.90)&0.923(0.94)&0.91(0.94)\\
$\nu3s_{1/2}$  & $0.24\pm0.03$  &0.31(0.30)&0.14(0.12)&0.09(0.07)&0.11(0.09)&$0.34\pm0.03$&0.41(0.41)&0.22(0.19)&0.16(0.14)&0.19(0.16)\\
$\nu2d_{3/2}$  & $0.18\pm0.025$ &0.31(0.29)&\emph{0.19}(\emph{0.16})&0.12(0.10)&0.15(0.12)&$0.37\pm0.04$&\emph{0.40}(\emph{0.39})&0.27(0.25)&0.20(0.17)&0.23(0.20)\\
$\nu1h_{11/2}$ &$$              &0.08(0.06)&0.06(0.05)&0.05(0.04)&0.04(0.03)&$0.25\pm0.07$&0.10(0.09)&0.09(0.07)&0.08(0.06)&0.05(0.04)\\
\hline\hline
\multicolumn{1}{c}{\multirow{2}{*}{Orbit}} & \multicolumn{5}{c}{$^{118}$Sn}  & \multicolumn{5}{c}{$^{124}$Sn}  \\ \cline{2-6} \cline{7-11}
       &      Exp.      &      PKA1      &      PKO3      &      PKDD     &DD-ME2 &      Exp.     &      PKA1      &      PKO3      &      PKDD   &DD-ME2   \\ \hline
$\nu2d_{5/2}$	 &$1.00\pm0.10$&\emph{0.91}(\emph{0.93})&0.86(0.89)&0.88(\emph{0.90})&\emph{0.92}(\emph{0.94})&$1.00\pm0.09$&\emph{0.96}(\emph{0.97})&\emph{0.94}(\emph{0.95})&\emph{0.95}(\emph{0.96})&\emph{0.96}(\emph{0.97})\\
$\nu1g_{7/2}$	 &$0.78\pm0.19$&\emph{0.81}(\emph{0.83})&\emph{0.93}(\emph{0.94})&\emph{0.95}(\emph{0.96})&\emph{0.96}(\emph{0.97})&$0.81\pm0.23$&\emph{0.93}(\emph{0.94})&\emph{0.96}(\emph{0.97})&\emph{0.97}(\emph{0.98})&\emph{0.98}(\emph{0.98})\\
$\nu3s_{1/2}$	&$0.80\pm0.10$&0.65(0.67)&0.45(0.45)&0.40(0.40)&1.00(1.00)&$0.95\pm0.10$&\emph{0.88}(\emph{0.90})&0.77(0.78)&0.73(0.74)&\emph{1.00}(\emph{1.00})\\
$\nu2d_{3/2}$	 &$0.60\pm0.08$&\emph{0.61}(\emph{0.62})&0.52(0.52)&0.46(0.46)&0.56(0.58)&$0.75\pm0.10$&\emph{0.85}(\emph{0.87})&\emph{0.81}(\emph{0.82})&\emph{0.77}(\emph{0.78})&\emph{0.86}(0.88)\\
$\nu1h_{11/2}$&$0.35\pm0.08$ &0.17(0.16)&0.20(0.18)&0.20(0.18)&0.13(0.10)&$0.38\pm0.12$&\emph{0.46}(\emph{0.44})&\emph{0.47}(\emph{0.46})&\emph{0.49}(\emph{0.48})&\emph{0.43}(\emph{0.42})\\
\end{tabular}
\end{ruledtabular}
\end{table*}

As another direct effect, the modification on the pairing interaction will substantially change the pairing contributions to the energy functional. Taking the even Pb isotopes as the representatives, Table \ref{tab:Pbpair} shows the pairing energies from the RHFB and RHB calculations with the selected effective parameters. For comparison the results calculated by RHB with NL1 and by non-relativistic HFB with Gogny D1S are also shown \cite{Gonzalez1996PLB}. It should be mentioned that in the former RHB-NL1 calculations the effective pairing interaction was taken as the Gogny force D1S with additional strength factor $\zeta = 1.15$ whereas for the latter Gogny calculations it kept as the original one \cite{Gonzalez1996PLB}.

As seen from Table \ref{tab:Pbpair} the pairing energies are remarkably reduced in the calculations of PKA1, PKO3 and PKDD with the optimized pairing force $V^{pp}_{opt}$, which leads to weaker pairing contributions than those from NL1 and Gogny calculations. In this work the pairing interaction is optimized as referred to the OESs of binding energies of the odd isotopes. While in Ref. \cite{Gonzalez1996PLB} the strength factor $\zeta = 1.15$ for NL1 is simply determined to reproduce the pairing energies of the Gogny calculations for Pb isotopes. Such deviations between the models may also originate from the inconsistence between the relativistic mean field and non-relativistic pairing potential as mentioned before. As referred to the pairing energies, the PKA1 and PKO3 calculations with original Gogny pairing force present slightly stronger pairing effects than the Gogny ones while PKDD, as well as DD-ME2, presents much weaker pairing effects. This discrepancy between the RHF and RMF models can be interpreted by the values of the non-relativistic-type effective mass \cite{Long2006PLB1, Jaminon1989PRC}. With the inclusion of Fock terms, PKA1 and PKO3 have a fairly large effective mass, close to the non-relativistic ones. Whereas in general the RMF models present smaller effective masses, e.g., by PKDD and DD-ME2. Globally the level densities determined by PKA1 and PKO3 are then larger than PKDD and DD-ME2 such that stronger pairing effects are obtained by the former ones with the same pairing interaction. This may also partially explain the reason why we have different pairing strength factor from the previous NL1 calculations.

\begin{table*}[htbp]
\caption{The pairing energy of Pb isotopes obtained by the CDF calculations with PKA1, PKO3 and PKDD. The results inside/outside the parenthesis are presented by the calculations taking the Gogny force D1S with/without pairing strength factor $\zeta$ as the effective pairing interaction. The corresponding results of RHB theory with NL1 with $\zeta=1.15$ and of HFB theory with Gogny force \cite{Gonzalez1996PLB} are also shown for comparison. }\label{tab:Pbpair}
\begin{ruledtabular}
\begin{tabular}{cccccc}
 A	&        PKA1     &      PKO3       &        PKDD        &  NL1   & Gogny  \\ \hline
196	&$-24.72(-17.76)$ &	$-24.54(-17.04)$&	$-17.47(-11.62)$ &        &        \\
198	&$-22.40(-16.03)$ &	$-22.40(-15.38)$&	$-15.56(-10.14)$ &        &        \\
200	&$-19.40(-13.73)$ &	$-19.52(-13.13)$&	$-13.14(-8.34 )$ &        &        \\
202	&$-15.65(-10.83)$ &	$-15.86(-10.28)$&	$-10.23(-6.26 )$ &$-14.49$&$-14.41$\\
204	&$-11.14(-7.34 )$ &	$-11.34(-6.85 )$&   $-6.85	(-3.97)$ &$-10.41$&$-10.49$\\
206	&$-5.90 (-3.50 )$ &	$-6.00 (-3.11 )$&   $-3.17	(-1.48)$ &$-5.57 $&$-5.74 $\\
208	&$0.00  (0.00)  $ & $  0.00(0.00  )$&   $0.00   (0.00)$  &$0.00  $&$0.00  $\\
210	&$-4.81 (-2.93 )$ &	$-6.89 (-4.30 )$&   $-5.34	(-3.66)$ &$-4.79 $&$-4.19 $\\
212	&$-8.58 (-5.04 )$ &	$-12.00(-7.67 )$&   $-9.61	(-6.56)$ &$-8.66 $&$-8.64 $\\
214	&$-11.64(-6.37 )$ &	$-16.04(-10.33)$&	$-12.99(-8.87)$  &$-11.77$&$-12.89$\\
\end{tabular}
\end{ruledtabular}
\end{table*}

\begin{figure}[htbp]
\ifpdf
\includegraphics[width=0.48\textwidth]{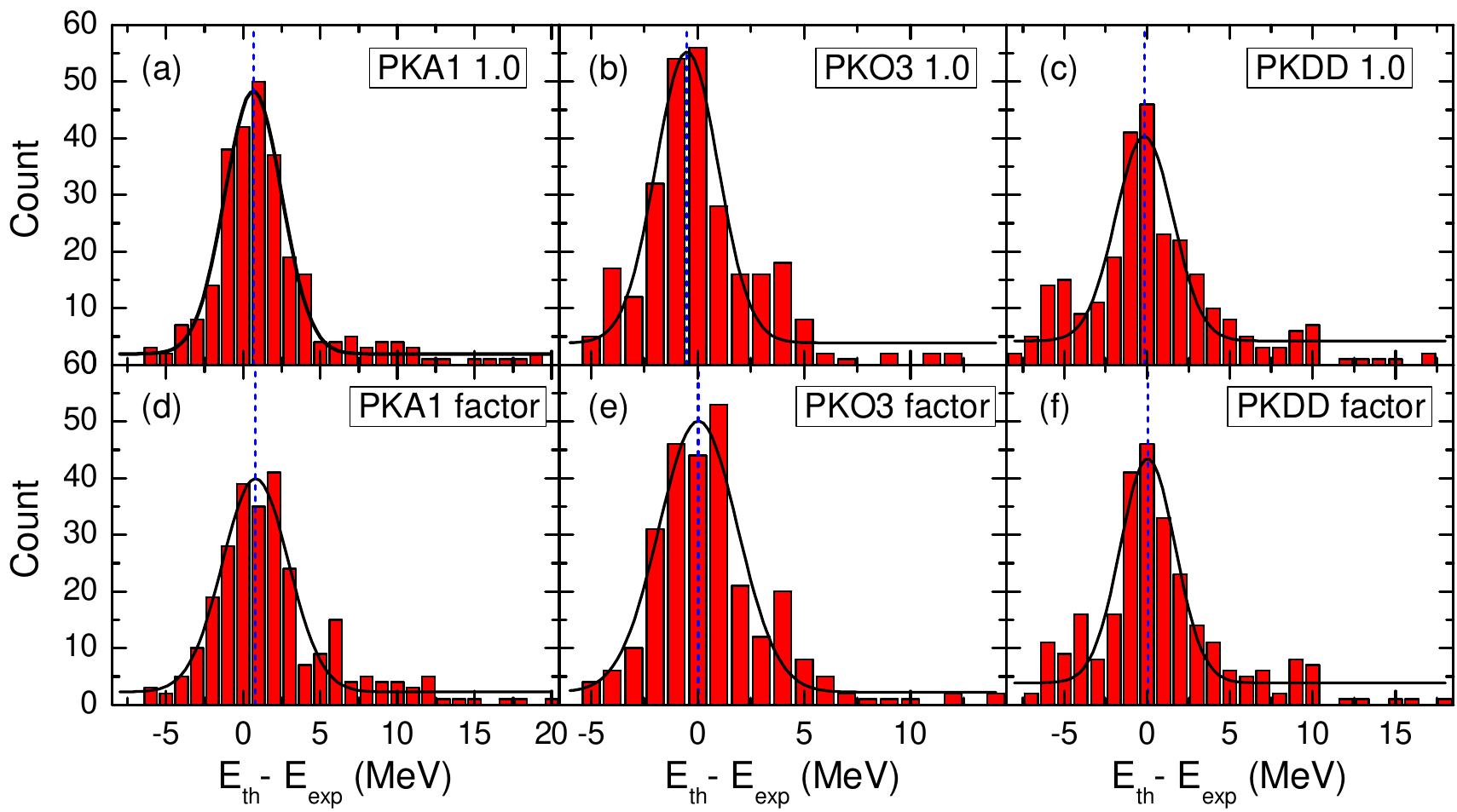}
\else
\includegraphics[width=0.48\textwidth]{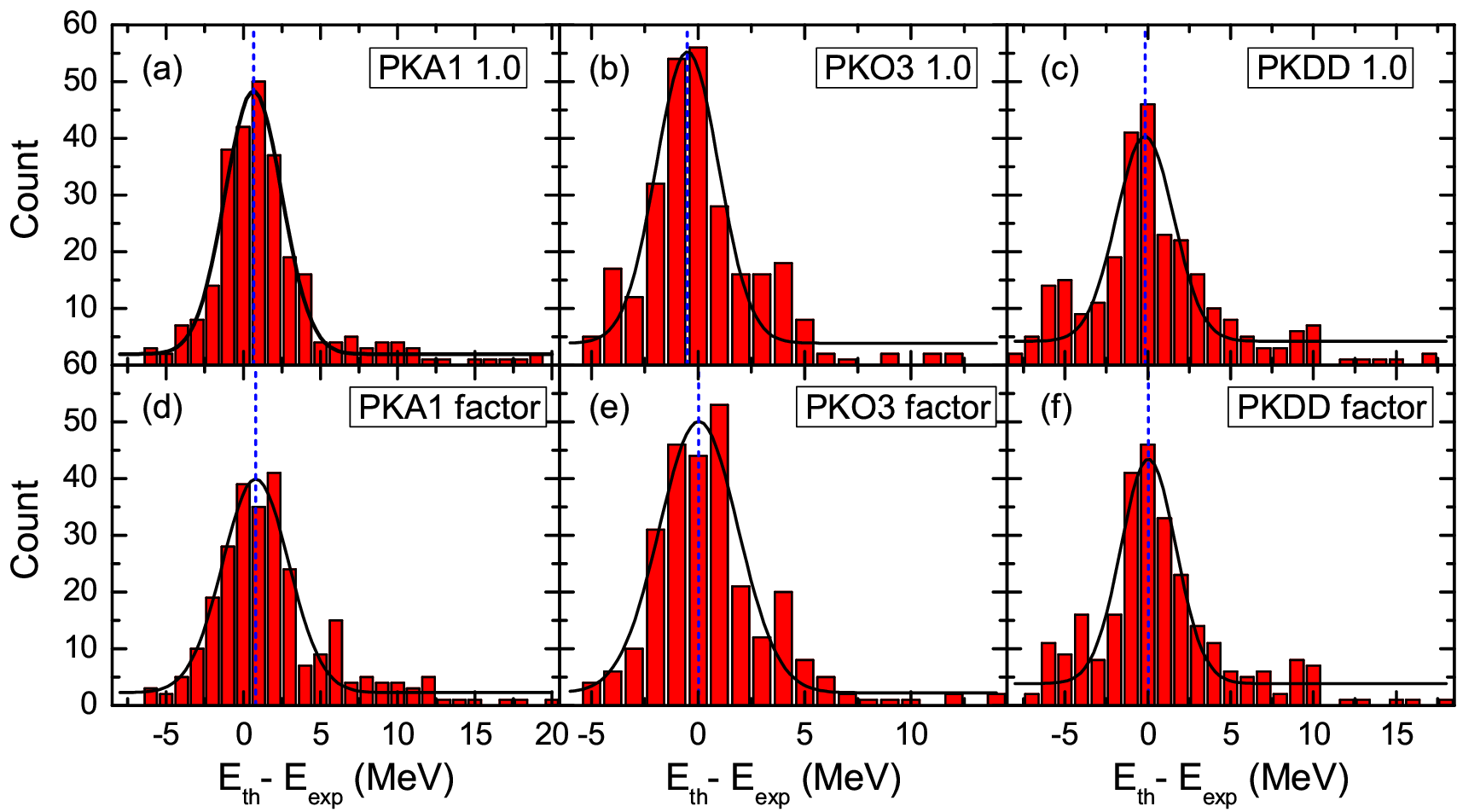}
\fi
\caption{\label{fig:fig6} (Color online) Distributions of deviations from experiment of the binding energies of 273 nuclei in the selected isotopes and isotones (1-MeV bins). Results obtained without (with) the effective pairing strength factor are shown in the upper (lower) panels. The best fitted Gaussian distributions [see Eq. (\ref{Gauss})] are also shown with black lines, and the corresponding parameters are listed in Table \ref{tab:tab3}.}
\end{figure}

\begin{figure}[htbp]
\ifpdf
\includegraphics[width=0.48\textwidth]{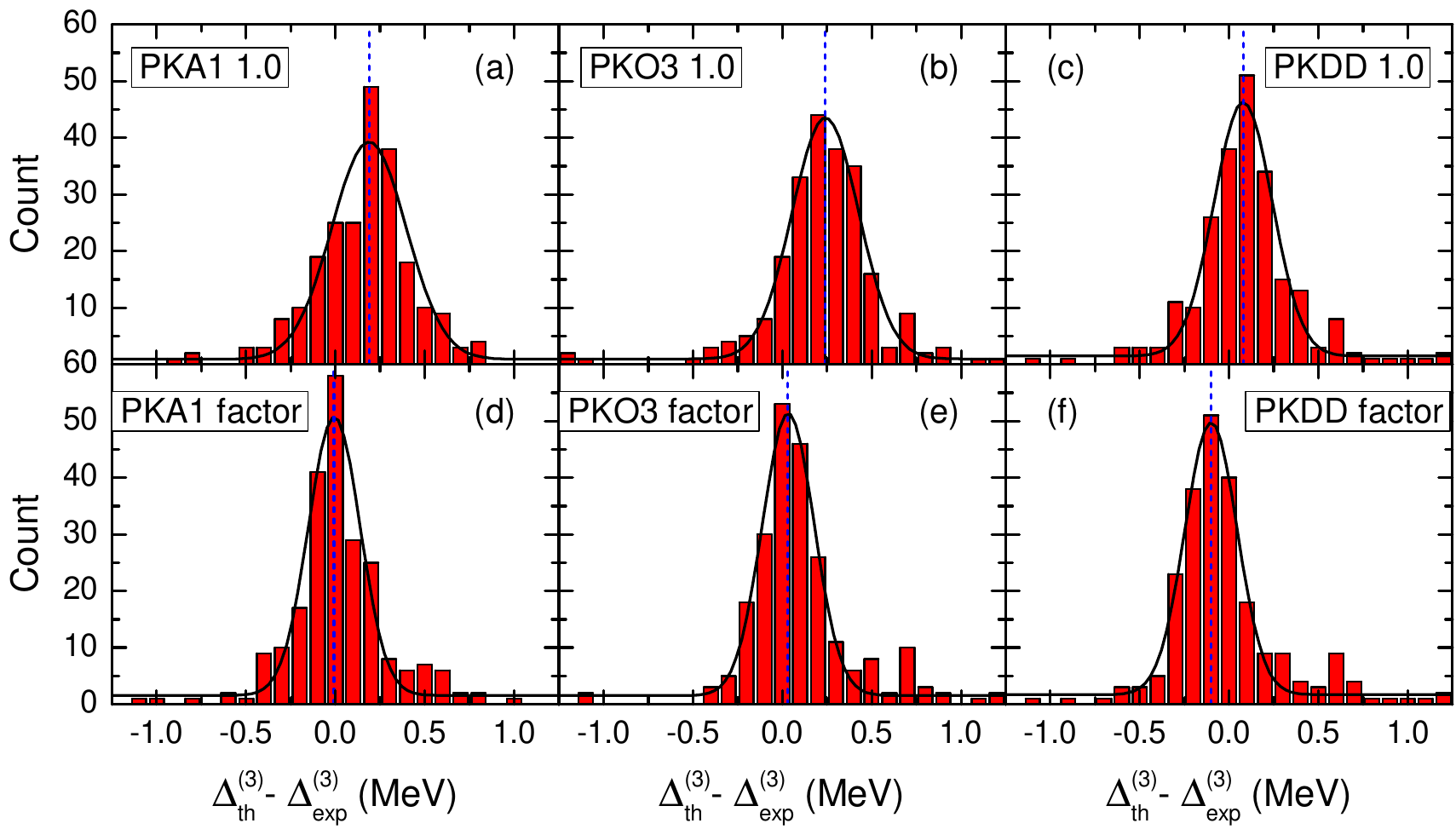}
\else
\includegraphics[width=0.48\textwidth]{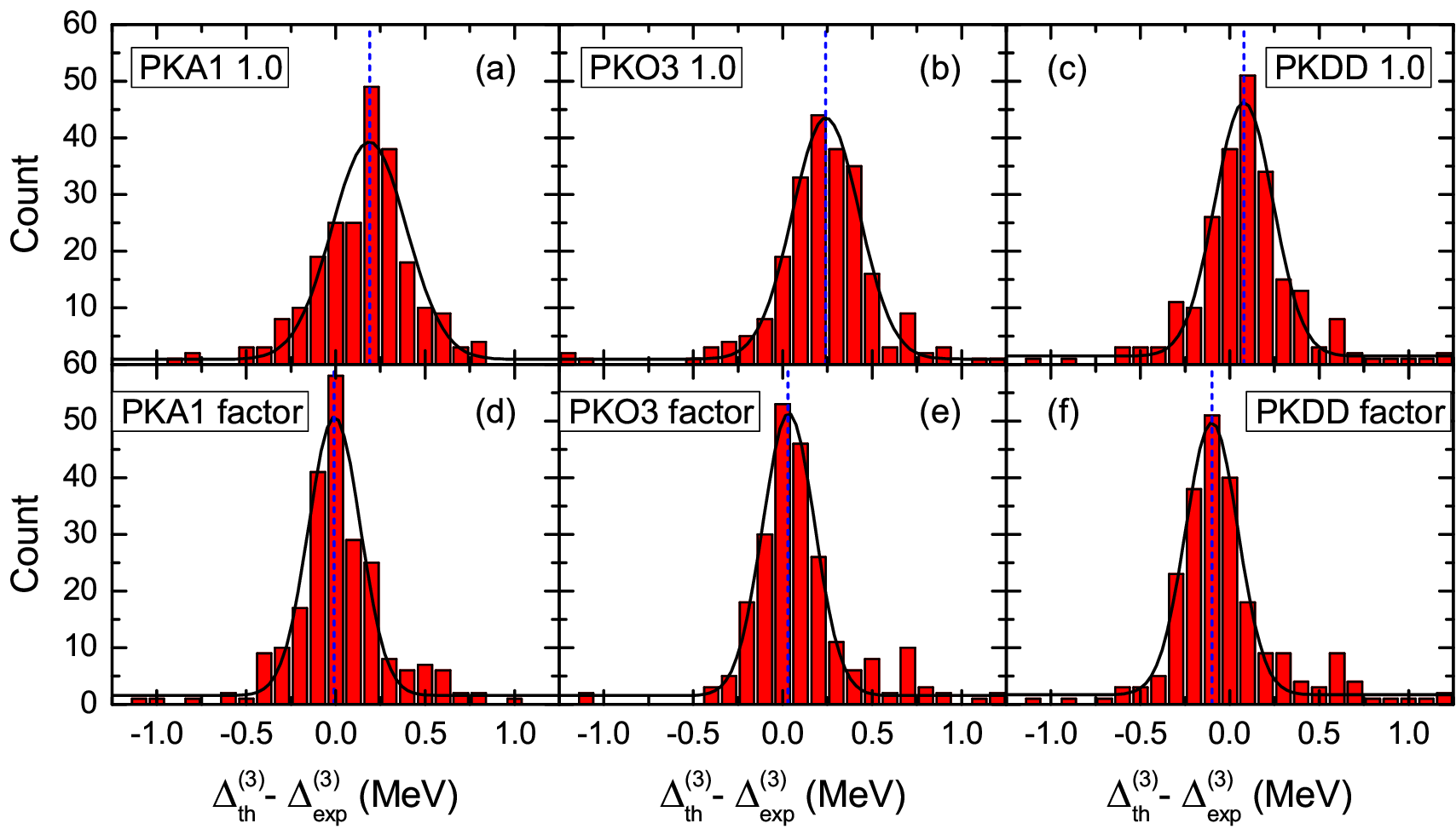}
\fi
\caption{\label{fig:fig7} (Color online) The same as Fig. \ref{fig:fig6} but for the OES [see Eq. (\ref{three point})] of 213 nuclei with both even and odd nucleon numbers in the selected isotopes and isotones (0.1-MeV bins).}
\end{figure}

\begin{table}[htbp]
\caption{\label{tab:tab3} Parameters of the corresponding Gaussian distributions [see Eq. (\ref{Gauss})] for the binding energy in Fig. \ref{fig:fig6} and the OES in Fig. \ref{fig:fig7}. Results obtained without/with the effective pairing strength factor are shown outside/inside the parenthesis. }
\begin{ruledtabular}
\begin{tabular}{ccccccccc}
	  & $y_0$        &     $x_c$          &     $w$           &     $A$        & \\ \hline
     & \multicolumn{4}{c}{Binding Energy}  \\ \cline{2-5}
PKA1 & 1.94(2.26)   & 0.66(0.80)         & 3.70(4.29)        & 214.64(202.45) \\
PKO3 & 3.88({2.23}) &-0.50(\emph{0.03})  & 2.94(3.71)        & 189.48(222.14) \\
PKDD & 4.23({3.86}) &-0.17(\emph{0.04})  & 3.46(\emph{3.32}) & 156.77(164.74) \\ \hline
     & \multicolumn{4}{c}{OES}     \\ \cline{2-5}
PKA1 &  0.89(1.57)  & 0.19(\emph{-0.01}) & 0.42(\emph{0.29}) & 20.05(17.82)    \\
PKO3 &  0.95(1.54)  & 0.24(\emph{ 0.03}) & 0.37(\emph{0.29}) & 19.96(18.13)    \\
PKDD &  1.48(1.70)  & 0.08(-0.10)        & 0.33(\emph{0.29}) & 18.32(17.62)    \\
\end{tabular}
\end{ruledtabular}
\end{table}

To provide a statistical understanding on the inconsistence between the non-relativistic Gogny pairing force and relativistic mean fields, Fig. \ref{fig:fig6} and Fig. \ref{fig:fig7} show the statistic behaviors of the deviations from the experimental data for both binding energies and OESs, respectively. For comparison, the results calculated with the original and optimized Gogny pairing forces are respectively shown in the upper and lower panels. In these two figures the solid curves denote the Guassian statistic fittings as,
\begin{align}\label{Gauss}
  y=y_0+\frac{A}{w\sqrt{\pi/2}} e^{-2\frac{(x-x_c)^2}{w^2}},
\end{align}
where $y_0$ is the minimum counts, $A$ represents the area of the Gaussian distribution, and $x_c$ and $w$ denote the statistic averages (ideally zero) and errors. In Fig. \ref{fig:fig6} and Fig. \ref{fig:fig7} it is clearly shown that the deviations of both binding energies and OESs from the data obey the Gaussian statistic behaviors. For the binding energy the deviations are not symmetrically distributed on the negative and positive sides and more counts are found for the positive deviations, mainly due to the deformation effects neglected in current calculations of Zr, Ce and Gd isotopes, as well as for the isotones. It is expected to get the results improved by including the deformation effects, which in general lead to more bound groud states. For the OESs the statistic distributions are nearly symmetric, especially for the PKA1 results with the optimized pairing force. Evident improvements due to the optimize pairing interaction can be also demonstrated from the statistic variables shown in Table \ref{tab:tab3}. It is seen that the statistic averages and errors on the OES deviations, i.e., the values of $x_c$ and $w$, are improved distinctly. While the improvements on the binding energies are not so distinct and systematical as the OESs. Especially for PKA1 the statistic qualities even become worse, inconsistent with the evident improvements on the OESs. This may also indicate the inconsistence between the relativistic mean field and non-relativistic pairing interaction.

\section{SUMMARY}\label{sum}

In summary, the neutron odd-even staggerings (OESs) of C, O, Ca, Ni, Zr, Sn, Ce, Gd and Pb isotopes, and the proton ones of $N=50$ and $82$ isotones, are studied systematically within the relativistic Hartree-Fock-Bogoliubov (RHFB) theory with PKA1 and PKO3, and the relativistic Hartree-Bogoliubov (RHB) theory with PKDD and DD-ME2. By taking the finite-range Gogny force D1S as an effective paring interaction, the neutron/proton OESs can be properly reproduced by the selected effective Lagragians.

Due to the inconsistence between the non-relativistic pairing force (Gogny D1S) and the relativistic effective Lagrangians (PKA1, PKO3, PKDD and DD-ME2), some systematic discrepancies with the data of the OESs are found, namely, the OESs and coherently the pairing effects are somewhat underestimated for light nuclei and overestimated for the heavy ones. Such inconsistence can be eliminated partially by introducing a $Z$- or $N$-dependent strength factor into the pairing force, which present better agreements with the experimental data of the OESs, by improving the root-mean-square deviation $\sigma_{\textrm{rms}}$ of the OES about $10\%$ for light nuclei and $30\%$ for heavy ones. Similar improvements are also obtained  on the description of single-particle configurations, especially in the calculations with PKA1. While with the optimization of pairing treatment, different even opposite systematical changes on the OESs and nuclear binding energies imply the remaining inconsistence between the relativistic mean fields and non-relativistic pairing interaction, which may call for an unified relativistic mechanism in both p-p and p-h channels as future perspectives.

\begin{acknowledgments}
This work is partly supported by the National Natural Science Foundation of China under Grant No. 11075066 and 11205075, the Fundamental Research Funds for the Central Universities under Contracts No. lzujbky-2012-k07 and No. lzujbky-2012-07, and the Program for New Century Excellent Talents in University.
\end{acknowledgments}

%%%%%%%%%%%%%%%%%%%%%%%%%%%%%%%%%%%%%%%%%%%%%%%%%%%%%%%%%%%%%%%%%%%%%%%%%%%%%%%%%%%%%%%%%%%%%%%%%%%%%%%%%%%%%%%%%%%%%%%%%%%%%%%%%%%%%%%%%%%%%%%%%%%%%%%%%%%

%\bibliographystyle{apsrev}	% (uses file "plain.bst"unsrt,abbrv,alpha,apsrev4-1)
%\bibliography{reference}	% expects file "myrefs.bib

\begin{thebibliography}{61}
\expandafter\ifx\csname natexlab\endcsname\relax\def\natexlab#1{#1}\fi
\expandafter\ifx\csname bibnamefont\endcsname\relax
  \def\bibnamefont#1{#1}\fi
\expandafter\ifx\csname bibfnamefont\endcsname\relax
  \def\bibfnamefont#1{#1}\fi
\expandafter\ifx\csname citenamefont\endcsname\relax
  \def\citenamefont#1{#1}\fi
\expandafter\ifx\csname url\endcsname\relax
  \def\url#1{\texttt{#1}}\fi
\expandafter\ifx\csname urlprefix\endcsname\relax\def\urlprefix{URL }\fi
\providecommand{\bibinfo}[2]{#2}
\providecommand{\eprint}[2][]{\url{#2}}

\bibitem[{\citenamefont{Bohr and Mottelson}(1998)}]{Bohr1998nuclear}
\bibinfo{author}{\bibfnamefont{A.}~\bibnamefont{Bohr}} \bibnamefont{and}
  \bibinfo{author}{\bibfnamefont{B.~R.} \bibnamefont{Mottelson}},
  \emph{\bibinfo{title}{Nuclear structure}}, vol.~\bibinfo{volume}{1}
  (\bibinfo{publisher}{World Scientific Publishing}, \bibinfo{year}{1998}).

\bibitem[{\citenamefont{Vretenar et~al.}(1998)\citenamefont{Vretenar,
  Lalazissis, and Ring}}]{Vretenar1997PRC}
\bibinfo{author}{\bibfnamefont{D.}~\bibnamefont{Vretenar}},
  \bibinfo{author}{\bibfnamefont{G.~A.} \bibnamefont{Lalazissis}},
  \bibnamefont{and} \bibinfo{author}{\bibfnamefont{P.}~\bibnamefont{Ring}},
  \bibinfo{journal}{Phys. Rev.} \textbf{\bibinfo{volume}{C 57}},
  \bibinfo{pages}{3071} (\bibinfo{year}{1998}).

\bibitem[{\citenamefont{Lalazissis and Raman}(1998)}]{Lalazissis1998PRC}
\bibinfo{author}{\bibfnamefont{G.~A.} \bibnamefont{Lalazissis}}
  \bibnamefont{and} \bibinfo{author}{\bibfnamefont{S.}~\bibnamefont{Raman}},
  \bibinfo{journal}{Phys. Rev.} \textbf{\bibinfo{volume}{C 58}},
  \bibinfo{pages}{1467} (\bibinfo{year}{1998}).

\bibitem[{\citenamefont{Lalazissis et~al.}(1998)\citenamefont{Lalazissis,
  Vretenar, P{\"o}schl, and Ring}}]{Lalazissis1998NPA}
\bibinfo{author}{\bibfnamefont{G.~A.} \bibnamefont{Lalazissis}},
  \bibinfo{author}{\bibfnamefont{D.}~\bibnamefont{Vretenar}},
  \bibinfo{author}{\bibfnamefont{W.}~\bibnamefont{P{\"o}schl}},
  \bibnamefont{and} \bibinfo{author}{\bibfnamefont{P.}~\bibnamefont{Ring}},
  \bibinfo{journal}{Nucl. Phys.} \textbf{\bibinfo{volume}{A 632}},
  \bibinfo{pages}{363} (\bibinfo{year}{1998}).

\bibitem[{\citenamefont{Meng and Ring}(1996)}]{Meng1996PRL}
\bibinfo{author}{\bibfnamefont{J.}~\bibnamefont{Meng}} \bibnamefont{and}
  \bibinfo{author}{\bibfnamefont{P.}~\bibnamefont{Ring}},
  \bibinfo{journal}{Phys. Rev. Lett.} \textbf{\bibinfo{volume}{77}},
  \bibinfo{pages}{3963} (\bibinfo{year}{1996}).

\bibitem[{\citenamefont{Meng and Ring}(1998)}]{Meng1998PRL}
\bibinfo{author}{\bibfnamefont{J.}~\bibnamefont{Meng}} \bibnamefont{and}
  \bibinfo{author}{\bibfnamefont{P.}~\bibnamefont{Ring}},
  \bibinfo{journal}{Phys. Rev. Lett.} \textbf{\bibinfo{volume}{80}},
  \bibinfo{pages}{460} (\bibinfo{year}{1998}).

\bibitem[{\citenamefont{Hagino et~al.}(2007)\citenamefont{Hagino, Sagawa,
  Carbonell, and Schuck}}]{Hagino2007PRL}
\bibinfo{author}{\bibfnamefont{K.}~\bibnamefont{Hagino}},
  \bibinfo{author}{\bibfnamefont{H.}~\bibnamefont{Sagawa}},
  \bibinfo{author}{\bibfnamefont{J.}~\bibnamefont{Carbonell}},
  \bibnamefont{and} \bibinfo{author}{\bibfnamefont{P.}~\bibnamefont{Schuck}},
  \bibinfo{journal}{Phys. Rev. Lett.} \textbf{\bibinfo{volume}{99}},
  \bibinfo{pages}{022506} (\bibinfo{year}{2007}).

\bibitem[{\citenamefont{Sun et~al.}(2010)\citenamefont{Sun, Toki, and
  Meng}}]{Sun2010PLB}
\bibinfo{author}{\bibfnamefont{B.~Y.} \bibnamefont{Sun}},
  \bibinfo{author}{\bibfnamefont{H.}~\bibnamefont{Toki}}, \bibnamefont{and}
  \bibinfo{author}{\bibfnamefont{J.}~\bibnamefont{Meng}},
  \bibinfo{journal}{Phys. Lett.} \textbf{\bibinfo{volume}{B 683}},
  \bibinfo{pages}{134} (\bibinfo{year}{2010}).

\bibitem[{\citenamefont{Sun et~al.}(2012)\citenamefont{Sun, Sun, and
  Meng}}]{Sun2012PRC}
\bibinfo{author}{\bibfnamefont{T.~T.} \bibnamefont{Sun}},
  \bibinfo{author}{\bibfnamefont{B.~Y.} \bibnamefont{Sun}}, \bibnamefont{and}
  \bibinfo{author}{\bibfnamefont{J.}~\bibnamefont{Meng}},
  \bibinfo{journal}{Phys. Rev.} \textbf{\bibinfo{volume}{C 86}},
  \bibinfo{pages}{014305} (\bibinfo{year}{2012}).

\bibitem[{\citenamefont{Sobiczewski and Pomorski}(2007)}]{Sobiczewski2007PPNP}
\bibinfo{author}{\bibfnamefont{A.}~\bibnamefont{Sobiczewski}} \bibnamefont{and}
  \bibinfo{author}{\bibfnamefont{K.}~\bibnamefont{Pomorski}},
  \bibinfo{journal}{Prog. Part. Nucl. Phys.} \textbf{\bibinfo{volume}{58}},
  \bibinfo{pages}{292} (\bibinfo{year}{2007}).

\bibitem[{\citenamefont{Karatzikos et~al.}(2010)\citenamefont{Karatzikos,
  Afanasjev, Lalazissis, and Ring}}]{Karatzikos2010PLB}
\bibinfo{author}{\bibfnamefont{S.}~\bibnamefont{Karatzikos}},
  \bibinfo{author}{\bibfnamefont{A.~V.} \bibnamefont{Afanasjev}},
  \bibinfo{author}{\bibfnamefont{G.~A.} \bibnamefont{Lalazissis}},
  \bibnamefont{and} \bibinfo{author}{\bibfnamefont{P.}~\bibnamefont{Ring}},
  \bibinfo{journal}{Phys. Lett.} \textbf{\bibinfo{volume}{B 689}},
  \bibinfo{pages}{72} (\bibinfo{year}{2010}).

\bibitem[{\citenamefont{Lalazissis et~al.}(1996)\citenamefont{Lalazissis,
  Sharma, Ring, and Gambhir}}]{Lalazissis1996NPA}
\bibinfo{author}{\bibfnamefont{G.~A.} \bibnamefont{Lalazissis}},
  \bibinfo{author}{\bibfnamefont{M.~M.} \bibnamefont{Sharma}},
  \bibinfo{author}{\bibfnamefont{P.}~\bibnamefont{Ring}}, \bibnamefont{and}
  \bibinfo{author}{\bibfnamefont{Y.~K.} \bibnamefont{Gambhir}},
  \bibinfo{journal}{Nucl. Phys.} \textbf{\bibinfo{volume}{A 608}},
  \bibinfo{pages}{202} (\bibinfo{year}{1996}).

\bibitem[{\citenamefont{Casten and Sherrill}(2000)}]{ExoticReview2000PPNP}
\bibinfo{author}{\bibfnamefont{R.~F.} \bibnamefont{Casten}} \bibnamefont{and}
  \bibinfo{author}{\bibfnamefont{B.~M.} \bibnamefont{Sherrill}},
  \bibinfo{journal}{Prog. Part. Nucl. Phys.} \textbf{\bibinfo{volume}{45}},
  \bibinfo{pages}{S171} (\bibinfo{year}{2000}).

\bibitem[{\citenamefont{Litvinov et~al.}(2005)}]{Litvinov2005PRL}
\bibinfo{author}{\bibfnamefont{Y.~A.} \bibnamefont{Litvinov}}
  \bibnamefont{et~al.}, \bibinfo{journal}{Phys. Rev. Lett.}
  \textbf{\bibinfo{volume}{95}}, \bibinfo{pages}{042501}
  (\bibinfo{year}{2005}).

\bibitem[{\citenamefont{Reinhard}(1989)}]{Reinhard1989}
\bibinfo{author}{\bibfnamefont{P.-G.} \bibnamefont{Reinhard}},
  \bibinfo{journal}{Rep. Prog. Phys} \textbf{\bibinfo{volume}{52}},
  \bibinfo{pages}{439} (\bibinfo{year}{1989}).

\bibitem[{\citenamefont{Ring}(1996)}]{Ring1996PPNP}
\bibinfo{author}{\bibfnamefont{P.}~\bibnamefont{Ring}}, \bibinfo{journal}{Prog.
  Part. Nucl. Phys.} \textbf{\bibinfo{volume}{37}}, \bibinfo{pages}{193}
  (\bibinfo{year}{1996}).

\bibitem[{\citenamefont{Bender et~al.}(2003)\citenamefont{Bender, Heenen, and
  Reinhard}}]{Bender2003RMP}
\bibinfo{author}{\bibfnamefont{M.}~\bibnamefont{Bender}},
  \bibinfo{author}{\bibfnamefont{P.-H.} \bibnamefont{Heenen}},
  \bibnamefont{and} \bibinfo{author}{\bibfnamefont{P.-G.}
  \bibnamefont{Reinhard}}, \bibinfo{journal}{Rev. Mod. Phys}
  \textbf{\bibinfo{volume}{75}}, \bibinfo{pages}{121} (\bibinfo{year}{2003}).

\bibitem[{\citenamefont{Vretenar et~al.}(2005)\citenamefont{Vretenar,
  Afanasjev, Lalazissis, and Ring}}]{Vretenar2005PhysRep}
\bibinfo{author}{\bibfnamefont{D.}~\bibnamefont{Vretenar}},
  \bibinfo{author}{\bibfnamefont{A.~V.} \bibnamefont{Afanasjev}},
  \bibinfo{author}{\bibfnamefont{G.~A.} \bibnamefont{Lalazissis}},
  \bibnamefont{and} \bibinfo{author}{\bibfnamefont{P.}~\bibnamefont{Ring}},
  \bibinfo{journal}{Phys. Rep.} \textbf{\bibinfo{volume}{409}},
  \bibinfo{pages}{101} (\bibinfo{year}{2005}).

\bibitem[{\citenamefont{Meng et~al.}(2006)\citenamefont{Meng, Toki, Zhou,
  Zhang, Long, and Geng}}]{Meng2006PPNP}
\bibinfo{author}{\bibfnamefont{J.}~\bibnamefont{Meng}},
  \bibinfo{author}{\bibfnamefont{H.}~\bibnamefont{Toki}},
  \bibinfo{author}{\bibfnamefont{S.~G.} \bibnamefont{Zhou}},
  \bibinfo{author}{\bibfnamefont{S.~Q.} \bibnamefont{Zhang}},
  \bibinfo{author}{\bibfnamefont{W.~H.} \bibnamefont{Long}}, \bibnamefont{and}
  \bibinfo{author}{\bibfnamefont{L.~S.} \bibnamefont{Geng}},
  \bibinfo{journal}{Prog. Part. Nucl. Phys.} \textbf{\bibinfo{volume}{57}},
  \bibinfo{pages}{470} (\bibinfo{year}{2006}).

\bibitem[{\citenamefont{Long et~al.}(2006{\natexlab{a}})\citenamefont{Long,
  Van~Giai, and Meng}}]{Long2006PLB1}
\bibinfo{author}{\bibfnamefont{W.~H.} \bibnamefont{Long}},
  \bibinfo{author}{\bibfnamefont{N.}~\bibnamefont{Van~Giai}}, \bibnamefont{and}
  \bibinfo{author}{\bibfnamefont{J.}~\bibnamefont{Meng}},
  \bibinfo{journal}{Phys. Lett.} \textbf{\bibinfo{volume}{B 640}},
  \bibinfo{pages}{150} (\bibinfo{year}{2006}{\natexlab{a}}).

\bibitem[{\citenamefont{Long et~al.}(2007)\citenamefont{Long, Sagawa, Van~Giai,
  and Meng}}]{Long2007PRC}
\bibinfo{author}{\bibfnamefont{W.~H.} \bibnamefont{Long}},
  \bibinfo{author}{\bibfnamefont{H.}~\bibnamefont{Sagawa}},
  \bibinfo{author}{\bibfnamefont{N.}~\bibnamefont{Van~Giai}}, \bibnamefont{and}
  \bibinfo{author}{\bibfnamefont{J.}~\bibnamefont{Meng}},
  \bibinfo{journal}{Phys. Rev.} \textbf{\bibinfo{volume}{C 76}},
  \bibinfo{pages}{034314} (\bibinfo{year}{2007}).

\bibitem[{\citenamefont{Walecka}(1974)}]{Walecka:1974}
\bibinfo{author}{\bibfnamefont{J.~D.} \bibnamefont{Walecka}},
  \bibinfo{journal}{Ann. Phys. (NY)} \textbf{\bibinfo{volume}{83}},
  \bibinfo{pages}{491} (\bibinfo{year}{1974}).

\bibitem[{\citenamefont{Serot and Walecka}(1986)}]{Serot:1986}
\bibinfo{author}{\bibfnamefont{B.~D.} \bibnamefont{Serot}} \bibnamefont{and}
  \bibinfo{author}{\bibfnamefont{J.~D.} \bibnamefont{Walecka}},
  \bibinfo{journal}{Adv. Nucl. Phys.} \textbf{\bibinfo{volume}{16}},
  \bibinfo{pages}{1} (\bibinfo{year}{1986}).

\bibitem[{\citenamefont{Paar et~al.}(2007)\citenamefont{Paar, Vretenar, Khan,
  and Col{\`o}}}]{Paar2007RPP}
\bibinfo{author}{\bibfnamefont{N.}~\bibnamefont{Paar}},
  \bibinfo{author}{\bibfnamefont{D.}~\bibnamefont{Vretenar}},
  \bibinfo{author}{\bibfnamefont{E.}~\bibnamefont{Khan}}, \bibnamefont{and}
  \bibinfo{author}{\bibfnamefont{G.}~\bibnamefont{Col{\`o}}},
  \bibinfo{journal}{Rep. Prog. Phys.} \textbf{\bibinfo{volume}{70}},
  \bibinfo{pages}{691} (\bibinfo{year}{2007}).

\bibitem[{\citenamefont{Lalazissis et~al.}(2009)\citenamefont{Lalazissis,
  Karatzikos, Serra, Otsuka, and Ring}}]{Lalazissis2009PRCR}
\bibinfo{author}{\bibfnamefont{G.~A.} \bibnamefont{Lalazissis}},
  \bibinfo{author}{\bibfnamefont{S.}~\bibnamefont{Karatzikos}},
  \bibinfo{author}{\bibfnamefont{M.}~\bibnamefont{Serra}},
  \bibinfo{author}{\bibfnamefont{T.}~\bibnamefont{Otsuka}}, \bibnamefont{and}
  \bibinfo{author}{\bibfnamefont{P.}~\bibnamefont{Ring}},
  \bibinfo{journal}{Phys. Rev.} \textbf{\bibinfo{volume}{C 80}},
  \bibinfo{pages}{041301} (\bibinfo{year}{2009}).

\bibitem[{\citenamefont{Long et~al.}(2010{\natexlab{a}})\citenamefont{Long,
  Ring, Van~Giai, and Meng}}]{Long2010PRC}
\bibinfo{author}{\bibfnamefont{W.~H.} \bibnamefont{Long}},
  \bibinfo{author}{\bibfnamefont{P.}~\bibnamefont{Ring}},
  \bibinfo{author}{\bibfnamefont{N.}~\bibnamefont{Van~Giai}}, \bibnamefont{and}
  \bibinfo{author}{\bibfnamefont{J.}~\bibnamefont{Meng}},
  \bibinfo{journal}{Phys. Rev.} \textbf{\bibinfo{volume}{C 81}},
  \bibinfo{pages}{024308} (\bibinfo{year}{2010}{\natexlab{a}}).

\bibitem[{\citenamefont{Long et~al.}(2008)\citenamefont{Long, Sagawa, Meng, and
  Van~Giai}}]{Long2008EPL}
\bibinfo{author}{\bibfnamefont{W.~H.} \bibnamefont{Long}},
  \bibinfo{author}{\bibfnamefont{H.}~\bibnamefont{Sagawa}},
  \bibinfo{author}{\bibfnamefont{J.}~\bibnamefont{Meng}}, \bibnamefont{and}
  \bibinfo{author}{\bibfnamefont{N.}~\bibnamefont{Van~Giai}},
  \bibinfo{journal}{Europhys Lett} \textbf{\bibinfo{volume}{82}},
  \bibinfo{pages}{12001} (\bibinfo{year}{2008}).

\bibitem[{\citenamefont{Long et~al.}(2009)\citenamefont{Long, Nakatsukasa,
  Sagawa, Meng, Nakada, and Zhang}}]{Long2009PLB}
\bibinfo{author}{\bibfnamefont{W.~H.} \bibnamefont{Long}},
  \bibinfo{author}{\bibfnamefont{T.}~\bibnamefont{Nakatsukasa}},
  \bibinfo{author}{\bibfnamefont{H.}~\bibnamefont{Sagawa}},
  \bibinfo{author}{\bibfnamefont{J.}~\bibnamefont{Meng}},
  \bibinfo{author}{\bibfnamefont{H.}~\bibnamefont{Nakada}}, \bibnamefont{and}
  \bibinfo{author}{\bibfnamefont{Y.}~\bibnamefont{Zhang}},
  \bibinfo{journal}{Phys. Lett.} \textbf{\bibinfo{volume}{B 680}},
  \bibinfo{pages}{428} (\bibinfo{year}{2009}).

\bibitem[{\citenamefont{Long et~al.}(2006{\natexlab{b}})\citenamefont{Long,
  Sagawa, Meng, and Van~Giai}}]{Long2006PLB2}
\bibinfo{author}{\bibfnamefont{W.~H.} \bibnamefont{Long}},
  \bibinfo{author}{\bibfnamefont{H.}~\bibnamefont{Sagawa}},
  \bibinfo{author}{\bibfnamefont{J.}~\bibnamefont{Meng}}, \bibnamefont{and}
  \bibinfo{author}{\bibfnamefont{N.}~\bibnamefont{Van~Giai}},
  \bibinfo{journal}{Phys. Lett.} \textbf{\bibinfo{volume}{B 639}},
  \bibinfo{pages}{242} (\bibinfo{year}{2006}{\natexlab{b}}).

\bibitem[{\citenamefont{Liang et~al.}(2010)\citenamefont{Liang, Long, Meng, and
  Giai}}]{Liang2010EPJA}
\bibinfo{author}{\bibfnamefont{H.}~\bibnamefont{Liang}},
  \bibinfo{author}{\bibfnamefont{W.~H.} \bibnamefont{Long}},
  \bibinfo{author}{\bibfnamefont{J.}~\bibnamefont{Meng}}, \bibnamefont{and}
  \bibinfo{author}{\bibfnamefont{N.~V.} \bibnamefont{Giai}},
  \bibinfo{journal}{Eur. Phys. J.} \textbf{\bibinfo{volume}{A 44}},
  \bibinfo{pages}{119} (\bibinfo{year}{2010}).

\bibitem[{\citenamefont{Long et~al.}(2010{\natexlab{b}})\citenamefont{Long,
  Ring, Meng, Van~Giai, and Bertulani}}]{Long2010PRCR}
\bibinfo{author}{\bibfnamefont{W.~H.} \bibnamefont{Long}},
  \bibinfo{author}{\bibfnamefont{P.}~\bibnamefont{Ring}},
  \bibinfo{author}{\bibfnamefont{J.}~\bibnamefont{Meng}},
  \bibinfo{author}{\bibfnamefont{N.}~\bibnamefont{Van~Giai}}, \bibnamefont{and}
  \bibinfo{author}{\bibfnamefont{C.~A.} \bibnamefont{Bertulani}},
  \bibinfo{journal}{Phys. Rev.} \textbf{\bibinfo{volume}{C 81}},
  \bibinfo{pages}{031302(R)} (\bibinfo{year}{2010}{\natexlab{b}}).

\bibitem[{\citenamefont{Liang et~al.}(2008)\citenamefont{Liang, Van~Giai, and
  Meng}}]{Liang2008PRL}
\bibinfo{author}{\bibfnamefont{H.}~\bibnamefont{Liang}},
  \bibinfo{author}{\bibfnamefont{N.}~\bibnamefont{Van~Giai}}, \bibnamefont{and}
  \bibinfo{author}{\bibfnamefont{J.}~\bibnamefont{Meng}},
  \bibinfo{journal}{Phys. Rev. Lett.} \textbf{\bibinfo{volume}{101}},
  \bibinfo{pages}{122502} (\bibinfo{year}{2008}).

\bibitem[{\citenamefont{Sun et~al.}(2008)\citenamefont{Sun, Long, Meng, and
  Lombardo}}]{Sun2008PRC}
\bibinfo{author}{\bibfnamefont{B.~Y.} \bibnamefont{Sun}},
  \bibinfo{author}{\bibfnamefont{W.~H.} \bibnamefont{Long}},
  \bibinfo{author}{\bibfnamefont{J.}~\bibnamefont{Meng}}, \bibnamefont{and}
  \bibinfo{author}{\bibfnamefont{U.}~\bibnamefont{Lombardo}},
  \bibinfo{journal}{Phys. Rev.} \textbf{\bibinfo{volume}{C 78}},
  \bibinfo{pages}{065805} (\bibinfo{year}{2008}).

\bibitem[{\citenamefont{Long et~al.}(2012)\citenamefont{Long, Sun, Hagino, and
  Sagawa}}]{Long2012PRC}
\bibinfo{author}{\bibfnamefont{W.~H.} \bibnamefont{Long}},
  \bibinfo{author}{\bibfnamefont{B.~Y.} \bibnamefont{Sun}},
  \bibinfo{author}{\bibfnamefont{K.}~\bibnamefont{Hagino}}, \bibnamefont{and}
  \bibinfo{author}{\bibfnamefont{H.}~\bibnamefont{Sagawa}},
  \bibinfo{journal}{Phys. Rev.} \textbf{\bibinfo{volume}{C 85}},
  \bibinfo{pages}{025806} (\bibinfo{year}{2012}).

\bibitem[{\citenamefont{Meng}(1998)}]{Meng1998NPA}
\bibinfo{author}{\bibfnamefont{J.}~\bibnamefont{Meng}}, \bibinfo{journal}{Nucl.
  Phys.} \textbf{\bibinfo{volume}{A 635}}, \bibinfo{pages}{3}
  (\bibinfo{year}{1998}).

\bibitem[{\citenamefont{Skyrme}(1956)}]{Skyrme1956PM}
\bibinfo{author}{\bibfnamefont{T.~H.~R.} \bibnamefont{Skyrme}},
  \bibinfo{journal}{Philos. Mag.} \textbf{\bibinfo{volume}{1}},
  \bibinfo{pages}{1043} (\bibinfo{year}{1956}).

\bibitem[{\citenamefont{Berger et~al.}(1991)\citenamefont{Berger, Girod, and
  Gogny}}]{Berger1991CPC}
\bibinfo{author}{\bibfnamefont{J.~F.} \bibnamefont{Berger}},
  \bibinfo{author}{\bibfnamefont{M.}~\bibnamefont{Girod}}, \bibnamefont{and}
  \bibinfo{author}{\bibfnamefont{D.}~\bibnamefont{Gogny}},
  \bibinfo{journal}{Comp. Phys. Comm.} \textbf{\bibinfo{volume}{63}},
  \bibinfo{pages}{365} (\bibinfo{year}{1991}).

\bibitem[{\citenamefont{Dobaczewski et~al.}(1996)\citenamefont{Dobaczewski,
  Nazarewicz, Werner, Berger, Chinn, and Decharg{\'e}}}]{Dobaczewski1996PRC}
\bibinfo{author}{\bibfnamefont{J.}~\bibnamefont{Dobaczewski}},
  \bibinfo{author}{\bibfnamefont{W.}~\bibnamefont{Nazarewicz}},
  \bibinfo{author}{\bibfnamefont{T.~R.} \bibnamefont{Werner}},
  \bibinfo{author}{\bibfnamefont{J.~F.} \bibnamefont{Berger}},
  \bibinfo{author}{\bibfnamefont{C.~R.} \bibnamefont{Chinn}}, \bibnamefont{and}
  \bibinfo{author}{\bibfnamefont{J.}~\bibnamefont{Decharg{\'e}}},
  \bibinfo{journal}{Phys. Rev. C} \textbf{\bibinfo{volume}{53}},
  \bibinfo{pages}{2809} (\bibinfo{year}{1996}).

\bibitem[{\citenamefont{Chappert et~al.}(2008)\citenamefont{Chappert, Girod,
  and Hilaire}}]{Chappert2008PLB}
\bibinfo{author}{\bibfnamefont{F.}~\bibnamefont{Chappert}},
  \bibinfo{author}{\bibfnamefont{M.}~\bibnamefont{Girod}}, \bibnamefont{and}
  \bibinfo{author}{\bibfnamefont{S.}~\bibnamefont{Hilaire}},
  \bibinfo{journal}{Phys. Lett.} \textbf{\bibinfo{volume}{B 668}},
  \bibinfo{pages}{420} (\bibinfo{year}{2008}).

\bibitem[{\citenamefont{Goriely et~al.}(2009)\citenamefont{Goriely, Hilaire,
  Girod, and P{\'e}ru}}]{Goriely2009PRL}
\bibinfo{author}{\bibfnamefont{S.}~\bibnamefont{Goriely}},
  \bibinfo{author}{\bibfnamefont{S.}~\bibnamefont{Hilaire}},
  \bibinfo{author}{\bibfnamefont{M.}~\bibnamefont{Girod}}, \bibnamefont{and}
  \bibinfo{author}{\bibfnamefont{S.}~\bibnamefont{P{\'e}ru}},
  \bibinfo{journal}{Phys. Rev. Lett.} \textbf{\bibinfo{volume}{102}},
  \bibinfo{pages}{242501} (\bibinfo{year}{2009}).

\bibitem[{\citenamefont{Gonzalez-Llarena
  et~al.}(1996)\citenamefont{Gonzalez-Llarena, Egido, Lalazissis, and
  Ring}}]{Gonzalez1996PLB}
\bibinfo{author}{\bibfnamefont{T.}~\bibnamefont{Gonzalez-Llarena}},
  \bibinfo{author}{\bibfnamefont{J.~L.} \bibnamefont{Egido}},
  \bibinfo{author}{\bibfnamefont{G.~A.} \bibnamefont{Lalazissis}},
  \bibnamefont{and} \bibinfo{author}{\bibfnamefont{P.}~\bibnamefont{Ring}},
  \bibinfo{journal}{Phys. Lett.} \textbf{\bibinfo{volume}{B 379}},
  \bibinfo{pages}{13} (\bibinfo{year}{1996}).

\bibitem[{\citenamefont{Afanasjev et~al.}(2003)\citenamefont{Afanasjev, Khoo,
  Frauendorf, Lalazissis, and Ahmad}}]{Afanasjev2003PRC}
\bibinfo{author}{\bibfnamefont{A.~V.} \bibnamefont{Afanasjev}},
  \bibinfo{author}{\bibfnamefont{T.~L.} \bibnamefont{Khoo}},
  \bibinfo{author}{\bibfnamefont{S.}~\bibnamefont{Frauendorf}},
  \bibinfo{author}{\bibfnamefont{G.~A.} \bibnamefont{Lalazissis}},
  \bibnamefont{and} \bibinfo{author}{\bibfnamefont{I.}~\bibnamefont{Ahmad}},
  \bibinfo{journal}{Phys. Rev.} \textbf{\bibinfo{volume}{C 67}},
  \bibinfo{pages}{024309} (\bibinfo{year}{2003}).

\bibitem[{\citenamefont{Bouyssy et~al.}(1987)\citenamefont{Bouyssy, Mathiot,
  Van~Giai, and Marcos}}]{Bouyssy1987PRC}
\bibinfo{author}{\bibfnamefont{A.}~\bibnamefont{Bouyssy}},
  \bibinfo{author}{\bibfnamefont{J.}~\bibnamefont{Mathiot}},
  \bibinfo{author}{\bibfnamefont{N.}~\bibnamefont{Van~Giai}}, \bibnamefont{and}
  \bibinfo{author}{\bibfnamefont{S.}~\bibnamefont{Marcos}},
  \bibinfo{journal}{Phys. Rev.} \textbf{\bibinfo{volume}{C 36}},
  \bibinfo{pages}{380} (\bibinfo{year}{1987}).

\bibitem[{\citenamefont{Gorkov}(1958)}]{Gorkov1958}
\bibinfo{author}{\bibfnamefont{L.~P.} \bibnamefont{Gorkov}},
  \bibinfo{journal}{Sov. Phys. JETP} \textbf{\bibinfo{volume}{7}},
  \bibinfo{pages}{505} (\bibinfo{year}{1958}).

\bibitem[{\citenamefont{Kucharek and Ring}(1991)}]{Kucharek1991ZPA}
\bibinfo{author}{\bibfnamefont{H.}~\bibnamefont{Kucharek}} \bibnamefont{and}
  \bibinfo{author}{\bibfnamefont{P.}~\bibnamefont{Ring}}, \bibinfo{journal}{Z.
  Phys.} \textbf{\bibinfo{volume}{A 339}}, \bibinfo{pages}{23}
  (\bibinfo{year}{1991}).

\bibitem[{\citenamefont{Zhou et~al.}(2003)\citenamefont{Zhou, Meng, and
  Ring}}]{Zhou2003PRC}
\bibinfo{author}{\bibfnamefont{S.~G.} \bibnamefont{Zhou}},
  \bibinfo{author}{\bibfnamefont{J.}~\bibnamefont{Meng}}, \bibnamefont{and}
  \bibinfo{author}{\bibfnamefont{P.}~\bibnamefont{Ring}},
  \bibinfo{journal}{Phys. Rev.} \textbf{\bibinfo{volume}{C 68}},
  \bibinfo{pages}{034323} (\bibinfo{year}{2003}).

\bibitem[{\citenamefont{Bender et~al.}(2000)\citenamefont{Bender, Rutz,
  Reinhard, and Maruhn}}]{Bender2000EPJA}
\bibinfo{author}{\bibfnamefont{M.}~\bibnamefont{Bender}},
  \bibinfo{author}{\bibfnamefont{K.}~\bibnamefont{Rutz}},
  \bibinfo{author}{\bibfnamefont{P.-G.} \bibnamefont{Reinhard}},
  \bibnamefont{and} \bibinfo{author}{\bibfnamefont{J.~A.}
  \bibnamefont{Maruhn}}, \bibinfo{journal}{Eur. Phys. J.}
  \textbf{\bibinfo{volume}{A 8}}, \bibinfo{pages}{59} (\bibinfo{year}{2000}).

\bibitem[{\citenamefont{Dobaczewski et~al.}(2001)\citenamefont{Dobaczewski,
  Magierski, Nazarewicz, Satula, and Szymanski}}]{Dobaczewski2001PRC}
\bibinfo{author}{\bibfnamefont{J.}~\bibnamefont{Dobaczewski}},
  \bibinfo{author}{\bibfnamefont{P.}~\bibnamefont{Magierski}},
  \bibinfo{author}{\bibfnamefont{W.}~\bibnamefont{Nazarewicz}},
  \bibinfo{author}{\bibfnamefont{W.}~\bibnamefont{Satula}}, \bibnamefont{and}
  \bibinfo{author}{\bibfnamefont{Z.}~\bibnamefont{Szymanski}},
  \bibinfo{journal}{Phys. Rev.} \textbf{\bibinfo{volume}{C 63}},
  \bibinfo{pages}{024308} (\bibinfo{year}{2001}).

\bibitem[{\citenamefont{Satu{\l}a et~al.}(1998)\citenamefont{Satu{\l}a,
  Dobaczewski, and Nazarewicz}}]{Satula1998PRL}
\bibinfo{author}{\bibfnamefont{W.}~\bibnamefont{Satu{\l}a}},
  \bibinfo{author}{\bibfnamefont{J.}~\bibnamefont{Dobaczewski}},
  \bibnamefont{and}
  \bibinfo{author}{\bibfnamefont{W.}~\bibnamefont{Nazarewicz}},
  \bibinfo{journal}{Phys. Rev. Lett.} \textbf{\bibinfo{volume}{81}},
  \bibinfo{pages}{3599} (\bibinfo{year}{1998}).

\bibitem[{\citenamefont{Bertsch et~al.}(2009)\citenamefont{Bertsch, Bertulani,
  Nazarewicz, Schunck, and Stoitsov}}]{Bertsch2009PRC}
\bibinfo{author}{\bibfnamefont{G.~F.} \bibnamefont{Bertsch}},
  \bibinfo{author}{\bibfnamefont{C.~A.} \bibnamefont{Bertulani}},
  \bibinfo{author}{\bibfnamefont{W.}~\bibnamefont{Nazarewicz}},
  \bibinfo{author}{\bibfnamefont{N.}~\bibnamefont{Schunck}}, \bibnamefont{and}
  \bibinfo{author}{\bibfnamefont{M.~V.} \bibnamefont{Stoitsov}},
  \bibinfo{journal}{Phys. Rev.} \textbf{\bibinfo{volume}{C 79}},
  \bibinfo{pages}{034306} (\bibinfo{year}{2009}).

\bibitem[{\citenamefont{Berger et~al.}(1984)\citenamefont{Berger, Girod, and
  Gogny}}]{Berger1984NPA}
\bibinfo{author}{\bibfnamefont{J.~F.} \bibnamefont{Berger}},
  \bibinfo{author}{\bibfnamefont{M.}~\bibnamefont{Girod}}, \bibnamefont{and}
  \bibinfo{author}{\bibfnamefont{D.}~\bibnamefont{Gogny}},
  \bibinfo{journal}{Nucl. Phys.} \textbf{\bibinfo{volume}{A 428}},
  \bibinfo{pages}{23} (\bibinfo{year}{1984}).

\bibitem[{\citenamefont{Long et~al.}(2004)\citenamefont{Long, Meng, Van~Giai,
  and Zhou}}]{Long2004PRC}
\bibinfo{author}{\bibfnamefont{W.~H.} \bibnamefont{Long}},
  \bibinfo{author}{\bibfnamefont{J.}~\bibnamefont{Meng}},
  \bibinfo{author}{\bibfnamefont{N.}~\bibnamefont{Van~Giai}}, \bibnamefont{and}
  \bibinfo{author}{\bibfnamefont{S.~G.} \bibnamefont{Zhou}},
  \bibinfo{journal}{Phys. Rev.} \textbf{\bibinfo{volume}{C 69}},
  \bibinfo{pages}{034319} (\bibinfo{year}{2004}).

\bibitem[{\citenamefont{Lalazissis et~al.}(2005)\citenamefont{Lalazissis,
  Nik\v{s}i\'{c}, Vretenar, and Ring}}]{Lalazissis2005PRC}
\bibinfo{author}{\bibfnamefont{G.~A.} \bibnamefont{Lalazissis}},
  \bibinfo{author}{\bibfnamefont{T.}~\bibnamefont{Nik\v{s}i\'{c}}},
  \bibinfo{author}{\bibfnamefont{D.}~\bibnamefont{Vretenar}}, \bibnamefont{and}
  \bibinfo{author}{\bibfnamefont{P.}~\bibnamefont{Ring}},
  \bibinfo{journal}{Phys. Rev.} \textbf{\bibinfo{volume}{C 71}},
  \bibinfo{pages}{024312} (\bibinfo{year}{2005}).

\bibitem[{\citenamefont{Afanasjev and Abusara}(2010)}]{Afanasjev2010PRC}
\bibinfo{author}{\bibfnamefont{A.~V.} \bibnamefont{Afanasjev}}
  \bibnamefont{and} \bibinfo{author}{\bibfnamefont{H.}~\bibnamefont{Abusara}},
  \bibinfo{journal}{Phys. Rev.} \textbf{\bibinfo{volume}{C 81}},
  \bibinfo{pages}{014309} (\bibinfo{year}{2010}).

\bibitem[{\citenamefont{Wapstra et~al.}(2003)\citenamefont{Wapstra, Audi, and
  Thibault}}]{Audi2003NPA}
\bibinfo{author}{\bibfnamefont{A.~H.} \bibnamefont{Wapstra}},
  \bibinfo{author}{\bibfnamefont{G.}~\bibnamefont{Audi}}, \bibnamefont{and}
  \bibinfo{author}{\bibfnamefont{C.}~\bibnamefont{Thibault}},
  \bibinfo{journal}{Nucl. Phys.} \textbf{\bibinfo{volume}{A 729}},
  \bibinfo{pages}{129} (\bibinfo{year}{2003}).

\bibitem[{\citenamefont{Moller et~al.}(1995)\citenamefont{Moller, Nix, Myers,
  and Swiatecki}}]{Moller1995ADNDT}
\bibinfo{author}{\bibfnamefont{P.}~\bibnamefont{Moller}},
  \bibinfo{author}{\bibfnamefont{J.~R.} \bibnamefont{Nix}},
  \bibinfo{author}{\bibfnamefont{W.~D.} \bibnamefont{Myers}}, \bibnamefont{and}
  \bibinfo{author}{\bibfnamefont{W.~J.} \bibnamefont{Swiatecki}},
  \bibinfo{journal}{At. Data Nucl. Data Tables} \textbf{\bibinfo{volume}{59}},
  \bibinfo{pages}{185} (\bibinfo{year}{1995}).

\bibitem[{\citenamefont{Wang et~al.}(2012)\citenamefont{Wang, Dong, and
  Long}}]{Wang2012tensor}
\bibinfo{author}{\bibfnamefont{L.~J.} \bibnamefont{Wang}},
  \bibinfo{author}{\bibfnamefont{J.~M.} \bibnamefont{Dong}}, \bibnamefont{and}
  \bibinfo{author}{\bibfnamefont{W.~H.} \bibnamefont{Long}},
  \bibinfo{journal}{arXiv:1210.5382}  (\bibinfo{year}{2012}).

\bibitem[{\citenamefont{Geng et~al.}(2006)\citenamefont{Geng, Meng, Toki, Long,
  and Shen}}]{Geng2006CPL}
\bibinfo{author}{\bibfnamefont{L.~S.} \bibnamefont{Geng}},
  \bibinfo{author}{\bibfnamefont{J.}~\bibnamefont{Meng}},
  \bibinfo{author}{\bibfnamefont{H.}~\bibnamefont{Toki}},
  \bibinfo{author}{\bibfnamefont{W.~H.} \bibnamefont{Long}}, \bibnamefont{and}
  \bibinfo{author}{\bibfnamefont{G.}~\bibnamefont{Shen}},
  \bibinfo{journal}{Chin. Phys. Lett} \textbf{\bibinfo{volume}{23}},
  \bibinfo{pages}{1139} (\bibinfo{year}{2006}).

\bibitem[{\citenamefont{Andreozzi et~al.}(1996)\citenamefont{Andreozzi,
  Coraggio, Covello, Gargano, and Porrino}}]{Andreozzi1996ZPA}
\bibinfo{author}{\bibfnamefont{F.}~\bibnamefont{Andreozzi}},
  \bibinfo{author}{\bibfnamefont{L.}~\bibnamefont{Coraggio}},
  \bibinfo{author}{\bibfnamefont{A.}~\bibnamefont{Covello}},
  \bibinfo{author}{\bibfnamefont{A.}~\bibnamefont{Gargano}}, \bibnamefont{and}
  \bibinfo{author}{\bibfnamefont{A.}~\bibnamefont{Porrino}},
  \bibinfo{journal}{Z. Phys.} \textbf{\bibinfo{volume}{A 354}},
  \bibinfo{pages}{253} (\bibinfo{year}{1996}).

\bibitem[{\citenamefont{Fleming}(1982)}]{Fleming}
\bibinfo{author}{\bibfnamefont{D.~G.} \bibnamefont{Fleming}},
  \bibinfo{journal}{J. Phys.} \textbf{\bibinfo{volume}{60}},
  \bibinfo{pages}{428} (\bibinfo{year}{1982}).

\bibitem[{\citenamefont{Jaminon and Mahaux}(1989)}]{Jaminon1989PRC}
\bibinfo{author}{\bibfnamefont{M.}~\bibnamefont{Jaminon}} \bibnamefont{and}
  \bibinfo{author}{\bibfnamefont{C.}~\bibnamefont{Mahaux}},
  \bibinfo{journal}{Phys. Rev.} \textbf{\bibinfo{volume}{C 40}},
  \bibinfo{pages}{354} (\bibinfo{year}{1989}).

\end{thebibliography}

\end{document}